\documentclass[nofootinbib,preprint,amsmath,amssymb,showpacs,showkeys,aps]{revtex4-1}
\usepackage{graphicx}
\usepackage[utf8,latin1]{inputenc}
\usepackage{dcolumn}
\usepackage{mathrsfs}
\usepackage{subcaption}
\usepackage{academicons}
\usepackage{mathtools, nccmath}
\usepackage{tensor}
\usepackage[overload]{empheq}
\usepackage{tikz,xcolor}
\usepackage{cases}
\usepackage{setspace}
\usepackage{lipsum}
\usepackage[normalem]{ulem}
\usepackage{bm}
\usepackage{soul}
\usepackage{hhline}
\usepackage{bigints}
\usepackage{hyperref}
\usepackage{amsmath}
\usepackage[T1]{fontenc}
\usepackage{fancyhdr}
\newcommand{\qm}[1]{``#1''}
\usepackage{hyperref}
\hypersetup{colorlinks, linkcolor={red},citecolor={blue},urlcolor={blue}}  

\newcommand{\bsubeqs}{\begin{subequations}}
\newcommand{\esubeqs}{\end{subequations}}

\definecolor{lime}{HTML}{A6CE39}
\DeclareRobustCommand{\orcidicon}{
	\begin{tikzpicture}
	\draw[lime, fill=lime] (0,0) 
	circle [radius=0.16] 
	node[white] {{\fontfamily{qag}\selectfont \tiny ID}};
	\draw[white, fill=white] (-0.0625,0.095) 
	circle [radius=0.007];
	\end{tikzpicture}
	\hspace{-2mm}
}

\newcommand{\dd}{{\rm d}}
\newcommand{\OO}{{\rm O}}
\newcommand{\OR}[1]{{\rm O}\left(r^{#1}\right)}
\newcommand{\RS}{R_{\rm S}}
\newcommand{\RPS}{\mathcal{R}_{\rm ps}}
\newcommand{\R}{\mathcal{R}}
\newcommand{\MP}{M_{\rm P}}
\newcommand{\LP}{\ell_{\rm P}}

\fancypagestyle{plain}{%
  \fancyhf{}
  \fancyfoot[C]{\iffloatpage{}{\thepage}}
  }
\foreach \x in {A, ..., Z}{%
	\expandafter\xdef\csname orcid\x\endcsname{\noexpand\href{https://orcid.org/\csname orcidauthor\x\endcsname}{\noexpand\orcidicon}}
}

\begin{document}

\title[Dynamical features and shadows of quantum Schwarzschild \\  black hole in effective field theories of gravity]{Dynamical features and shadows of quantum Schwarzschild \\ black hole in effective field theories of gravity}

\author{Zi-Liang Wang\orcidA{}$^{1}$}\email{ziliang.wang@just.edu.cn}
\author{Emmanuele Battista\orcidB{}$^{2,3,4}$}\email{ebattista@na.infn.it}\email{emmanuelebattista@gmail.com}

\affiliation{$^1$ Department of Physics, School of Science, Jiangsu University of Science and Technology, Zhenjiang, 212003, China
\\
$^2$ Quantum  Theory Center ($\hbar$QTC) \& D-IAS, Southern Denmark University, Campusvej 55, 5230 Odense M, Denmark
\\
$^3$ Department of Physics ``Ettore Pancini'', University of Naples ``Federico II'', Via Cintia Edificio 6, 80126 Naples, Italy\\
$^4$ Istituto Nazionale di Fisica Nucleare, Sezione di Napoli, University of Naples ``Federico II'', Via Cintia Edificio 6, 80126 Naples, Italy
 }

\date{\today}

\begin{abstract}

We investigate the properties of the Schwarzschild black hole geometry involving   leading one-loop long-distance quantum effects, which arise within the framework of effective field theories of gravity. Our analysis reveals  that  geodesic  trajectories of both massive and massless particles can assume completely different behaviours depending on the sign assumed by the quantum contributions, in spite of their smallness. Moreover,  we  find that the positions of stable and unstable circular orbits are determined by  an algebraic quartic equation, which we solve by developing a straightforward and analytic method. Additionally, we  examine black hole shadows and rings by means of two different emission profile models, which  account for quantum corrections to the innermost stable circular orbit and photon sphere radii. The Hawking temperature and the entropy of the black hole are also  derived. Finally, we draw our conclusions.

\end{abstract}

\maketitle

\section{Introduction}
\label{sec:intro}

Among the four fundamental forces of nature, gravity is the most intriguing. Indeed, despite having  been  the first to be studied,   it remains the only interaction for which we have yet to provide a definitive quantum description. The main troubles in this regard can be traced back to  the fact that general relativity is not (perturbatively) renormalizable in the traditional sense. This aspect can be accounted for in different ways. First of all, on dimensional grounds, a simple power counting argument shows that the Newton constant $G$ carries the dimension of the inverse Planck mass squared (similarly to the Fermi model of weak interactions), while it is known that a renormalizable theory possesses a coupling constant with positive mass dimension \cite{Peskin1995}. Moreover,   one-loop infinities of vacuum gravity  involve quadratic combinations of the Ricci tensor, which are thus not proportional to the original Einstein-Hilbert Lagrangian \cite{tHooft1974}. The situation gets worse at two-loop level, where ultraviolet divergencies comprise cubic invariants of the Riemann tensor, and hence cannot be canceled even if Einstein field equations are employed \cite{Goroff1985}. This state of affairs can be readily explained thanks to the theorem illustrated in Ref.  \cite{vanNieuwenhuizen1976}, which delineates the general form assumed by the leading $L$-loop divergencies of the quantum $S$ matrix for pure gravity in $d$ dimensions. The immediate consequence of this result is that the cancellation of ultraviolet infinities would require the introduction in the counter-Lagrangian of an infinite number of contributions, which  are proportional to arbitrarily high powers of the Riemann curvature tensor and its covariant derivatives, and hence not present in the Einstein-Hilbert  Lagrangian.

To address this unsatisfactory scenario, some competing quantum formulations of gravity have been put forward thus far. The most promising frameworks are loop quantum gravity, which represents a background-independent and nonperturbative approach  \cite{Rovelli2004,Ashtekar2017}, and string theory, which relies on both perturbative and nonperturbative techniques and has the ambitious scope of unifying all known fundamental   physics into one single paradigm \cite{Aharony1999,Becker2006,Blumenhagen2013}. Other research programs  worth mentioning are: twistor theory \cite{Penrose1977},  noncommutative geometry \cite{Connes2008,Steinacker2024}, Euclidean quantum gravity \cite{Esposito1992,Esposito1997}, causal dynamical triangulation \cite{Loll2019}, and casual set theory \cite{Surya2019}. However, despite these efforts, a comprehensive and viable quantum gravity model is still missing. One way to overcome this issue  consists in resorting to the tools of effective field theories (EFTs), which permit  treating general relativity as a well-behaved quantum theory at energies far below the Planck mass \cite{Donoghue1994,Donoghue1995c,Donoghue2012a,Donoghue2017a,Battista-book2017,Latosh2019,Burgess2020,Donoghue2022,Bjerrum-Bohr2024}. Gravity naturally adheres to the EFT pattern.  The main concept is that, after having integrated heavy particles out of the theory, the remaining massless (or, in general, very light) fields will be described by the most general effective Lagrangian
which is  invariant under the action of the diffeomorphism group. This Lagrangian function can then be
ordered in an energy expansion, where the leading low-energy interactions stem from the Einstein-Hilbert piece  and can be used in a full field-theoretic manner to make physical  predictions. Indeed, the Ricci scalar occurring in the Einstein-Hilbert action contains second-order derivatives of the metric, which become quadratic in the four-momentum  $p_\mu \sim {\rm i} \partial_\mu$ when translated  in momentum space. On the other hand, higher-order terms are at least quartic in $p_\mu$ and hence are suppressed at low energy. The key strength of the EFT scheme thus lies in the fact that it can be employed  even if the high-energy completion of quantum gravity remains unknown. Indeed,  the only residual effects of high-energy contributions appear, at low-energy scales, in the form of a shift of a few parameters in the Lagrangian whose (renormalized) values  can be determined experimentally.

By means of the EFT formalism, it has been possible to  derive the leading  one-loop  long-distance quantum corrections to the Newtonian gravitational potential \cite{Donoghue1994PRL,Donoghue1994,Khriplovich2002,Bjerrum-Bohr2002,dePaulaNetto2021,Frob2021}, as well as the principal known solutions of Einstein field equations  \cite{Donoghue2001,Bohr2003}, and subsequently many applications,  both at classical and quantum level, of these results have been considered in the literature \cite{Battista2014a,Battista2014b,Battista2015a,Battista2015b,Tartaglia2018,Ansari2022,Battista2017a,Kim2022,Battista2020c,Alshaery2020,Singh2021,Abouelmagd2023,Yamada2010b,Yamada2010,Ichita2010,Yamada2015,Asada2022,Jenkins2018,Liu2023,Mandal2022,Latosh2023a,Latosh2021a,Arbuzov2020a,Latosh2020b,Ali2024}. Motivated by  the enormous complexity of the calculations involved (see e.g. Refs. \cite{Latosh2022a,Latosh2024}), EFT techniques have been recently combined with on-shell unitarity methods \cite{Bjerrum-Bohr2013}, unveiling interesting outcomes in the field of scattering amplitudes and  their relation to classical gravity \cite{Bjerrum-Bohr2018,Bjerrum-Bohr2021,Damgaard2021,Bjerrum-Bohr2021c,Vanhove2022}.

Another topic which has garnered attention  concerns the study of the deformations and  quantum aspects of  black holes, drawing from either fundamental or extended gravity patterns,    EFT approaches, and different quantum gravity models  (see e.g. Refs. \cite{Dymnikova1992,Bonanno2000,Hayward2005,Kirilin2006,Gonzalez2015,Calmet2017,Bargueno2016,Ashtekar2018,Calmet2019,Nicolini2019,Platania2019,Faraoni2020,Ruiz2021,Calmet2021b,Eichhorn2022,DAlise2023,DelPiano2023,Beltran-Palau2022,Gong2023,Mitra2023,Damgaard2024,DelPiano2024,Lin2024,Khan2024,Bhar2024}). Indeed, such investigations can give insights into the interplay between general relativity and quantum mechanics, providing a unique quantum gravity testing ground; moreover, the examination of these geometries offers the possibility of sorting out the issue of classical spacetime singularities, furnish crucial clues to understanding the behavior of the early universe \cite{Hooper2019}, and give hints for resolving the black hole information paradox \cite{Maldacena2001}. 

Recently, we have delved into the quantum Schwarzschild solution  within the EFT pattern \cite{Battista2023}. Our study has been carried out upon constructing 
a coordinate transformation relating the harmonic coordinates originally employed in Ref. \cite{Bohr2003} and the Schwarzschild ones.  To take into account the discordant results existing in the literature (see e.g. Table 1 in Ref. \cite{Bargueno2016}), the one-loop quantum corrections have been written in their most general form in terms  of a factor $k_1$,  and we have found that the metric components are such that $-g_{tt} \neq g^{rr}$. This relation has led to several interesting implications not present at classical level, such as the possible occurrence of  a Penrose-like energy-extraction mechanism violating the null energy condition (NEC).  In this paper, we further explore the features of the quantum Schwarzschild geometry. Specifically, we  analyze the behaviour of both timelike and null geodesics, as well as the black hole appearance and its  emission properties. The plan of the paper is as follows. We begin with an outline of the main properties and the thermodynamic aspects of the quantum Schwarzschild solution in Sec. \ref{Sec:Quantum-metric}. After that,  the influence of the quantum effects on the dynamics of freely falling particles is examined in Sec. \ref{Sec:geodesic-motion-massive-massless}, where we also show that the positions of stable and unstable circular orbits are ruled by a quartic equation, for which we devise a simple and analytic   resolution method. The findings of this section are then exploited to deal with shadows and rings of the quantum black hole in Sec. \ref{Sec:Shadows-rings}. We finally draw our conclusions in Sec.  \ref{Sec:conclusions}.

\section{The quantum Schwarzschild geometry}\label{Sec:Quantum-metric}

Within the EFT paradigm, it is possible to work out the leading low-energy one-loop quantum corrections to the Schwarzschild metric \cite{Bohr2003}. The underlying calculations, which can be performed either by means of the traditional Feynman diagrammatic rules or the modern on-shell unitarity methods \cite{Bjerrum-Bohr2013},  involve harmonic coordinates. By constructing a coordinate transformation to the standard Schwarzschild coordinates $\{ct,r,\theta,\phi\}$, the quantum  
Schwarzschild metric can be written  as follows \cite{Battista2023}
\begin{subequations}
\label{Schwarzschild_metric_quantum-standard}
\begin{align}
{\rm d}s^2=g_{\mu \nu}{\rm d}x^\mu{\rm d}x^\nu= - B(r)\left(c^2{\rm d}t^2\right) + A(r){\rm d}r^2 + r^2 {\rm d}\Omega^2,
\end{align}
\text{with ${\rm d}\Omega^2 = {\rm d}\theta^2 + \sin^2 \theta \; {\rm d}\phi^2$ and} 
\begin{align} 
B(r) &= 1-\frac{\RS}{r} + \frac{k_1}{2} \frac{\RS \LP^2 }{ r^3} + \OO\left(r^{-4}\right),
\label{B-of-r-quantum}
\\
A(r) &= \left(1-\frac{\RS}{r} \right)^{-1} - \frac{3 k_1}{2} \frac{\RS \LP^2}{ r^3} + \OO\left(r^{-4}\right),
\label{A-of-r-quantum}
\end{align}
\end{subequations}
$R_{\rm S} = 2GM/c^2$ and $\ell_{\rm P}=\left(G \hbar/c^3\right)^{1/2}$ being  the Schwarzschild radius and  Planck length, respectively.  In the above equations, the remainder $\OO\left(r^{-4}\right)$ indicates two-loop quantum terms and will be hereafter omitted.  Moreover, $k_1$ is the dimensionless real-valued parameter embodying the quantum contributions;  since the literature provides different and sometimes discordant assessments, we will leave $k_1$ general; however, we note that its typical values are such that  (see e.g. Table 1 in Ref. \cite{Bargueno2016})
\begin{align}
\vert k_1 \vert \sim \OO (1).  
\label{k1-O1}
\end{align} 
It has been  demonstrated in Ref. \cite{Bjerrum-Bohr2013} that  the quantum contributions appearing in Eq. \eqref{Schwarzschild_metric_quantum-standard} have a  universal character, in  the sense that they are  to be expected  in any quantum gravity theory with the same low-energy degrees of freedom as those considered in this paper. 

As explained in Ref. \cite{Battista2023},  the validity of the EFT scheme requires that
\begin{align} \label{Rs-bigger-than-lp}
  R_{\rm S} \gg \ell_{\rm P},  
\end{align}
or, equivalently,
\begin{align} \label{M-bigger-M-Planck}
  M \gg \frac{1}{2}  M_{\rm P},
\end{align}
where $M_{\rm P} = \left(\hbar c/G\right)^{1/2}$ denotes  Planck mass. In fact, these relations guarantee that in our model the one-loop modifications to the standard classical outcomes
do not become arbitrarily large. In this way, our results can always be written as a classical piece amended by a small quantum factor.

One of the peculiarities of the metric \eqref{Schwarzschild_metric_quantum-standard} is  that  $-g_{tt} \neq g^{rr}$. This condition gives rise to some interesting features which will be discussed in Secs. \ref{Sec:horizon-null-hypers} and \ref{Sec:change-signature} (for further details, including the description of a possible energy-extraction mechanism, we refer the reader to Ref. \cite{Battista2023}). We conclude this section with an analysis of the effective energy-momentum tensor underlying the quantum Schwarzschild black hole and a first account of its thermodynamic properties  in   Secs. \ref{Sec:Effective-EMT}  and  \ref{Sec:thermod-aspects}, respectively.

\subsection{Metric horizons and null hypersurfaces}\label{Sec:horizon-null-hypers}

The horizons of the quantum spacetime \eqref{Schwarzschild_metric_quantum-standard} can be obtained from the condition $g_{tt}=0$. By solving the ensuing algebraic cubic equation  $r^3 -\RS r^2 + \tfrac{1}{2}k_1 \RS \LP^2=0$, we find that when $k_1$ is positive and $M > M^\star$ (with $M^\star :=  M_{\rm P} \, \sqrt{\tfrac{27}{32}k_1}$), the Schwarzschild metric admits two horizons whose radii are 
\begin{align}
r_1 &= \frac{2}{3}R_{\rm S} \cos \left\{ \frac{1}{3} \arccos \left[1-\frac{27}{4}k_1 \left(\frac{\ell_{\rm P}}{R_{\rm S}}\right)^2\right] \right\}   +\frac{1}{3} R_{\rm S}
\nonumber \\
&= R_{\rm S} \left[1-\frac{k_1}{2} \frac{\LP^2}{\RS^2} + \OO\left(\ell_{\rm P}^4/R_{\rm S}^4\right)\right],
\label{solution-r1}  
\\
r_2 &= -\frac{2}{3}R_{\rm S} \sin \left\{\frac{\pi}{6}- \frac{1}{3} \arccos \left[1-\frac{27}{4}k_1 \left(\frac{\ell_{\rm P}}{R_{\rm S}}\right)^2\right] \right\}   +\frac{1}{3} R_{\rm S}
\nonumber \\
&=\ell_{\rm P} \sqrt{\frac{k_1}{2}} + \OO\left(\ell_{\rm P}^2/R_{\rm S}\right).
\label{solution-r2}
\end{align}
On the other hand, when $k_1 <0$, then for any real-valued $M$ there exists one metric horizon only with  radius
\begin{align}
r_3&= \frac{1}{3} R_{\rm S} + \frac{1}{3}  R_{\rm S} \left[1-\frac{27}{4}k_1 \left(\frac{\ell_{\rm P}}{R_{\rm S}}\right)^2 + 3 \sqrt{3} \, \sqrt{\frac{k_1}{2}\left(\frac{\ell_{\rm P}}{R_{\rm S}}\right)^2 \left(k_1\frac{27}{8} \frac{\ell_{\rm P}^2}{R_{\rm S}^2}-1\right)}\right]^{-1/3}  
\nonumber \\
&+ \frac{1}{3}  R_{\rm S} \left[1-\frac{27}{4}k_1 \left(\frac{\ell_{\rm P}}{R_{\rm S}}\right)^2 + 3 \sqrt{3} \, \sqrt{\frac{k_1}{2}\left(\frac{\ell_{\rm P}}{R_{\rm S}}\right)^2 \left(k_1\frac{27}{8} \frac{\ell_{\rm P}^2}{R_{\rm S}^2}-1\right)}\right]^{1/3}  
\nonumber \\
&= R_{\rm S} \left[ 1-\frac{k_1}{2} \frac{\LP^2}{\RS^2} +\OO\left(\ell_{\rm P}^4/R_{\rm S}^4\right) \right].
\label{solution-r3}
\end{align}
It is worth noting that Eqs. \eqref{solution-r1}--\eqref{solution-r3} have been obtained by exploiting the constraint \eqref{Rs-bigger-than-lp}; moreover, as a consequence of  Eqs. \eqref{k1-O1} and \eqref{Rs-bigger-than-lp},  the horizons \eqref{solution-r1} and \eqref{solution-r2} meet the inequality 
\begin{align}
    r_2 \ll r_1.
\end{align}

\begin{figure}[bht!]
\centering\includegraphics[scale=0.30]{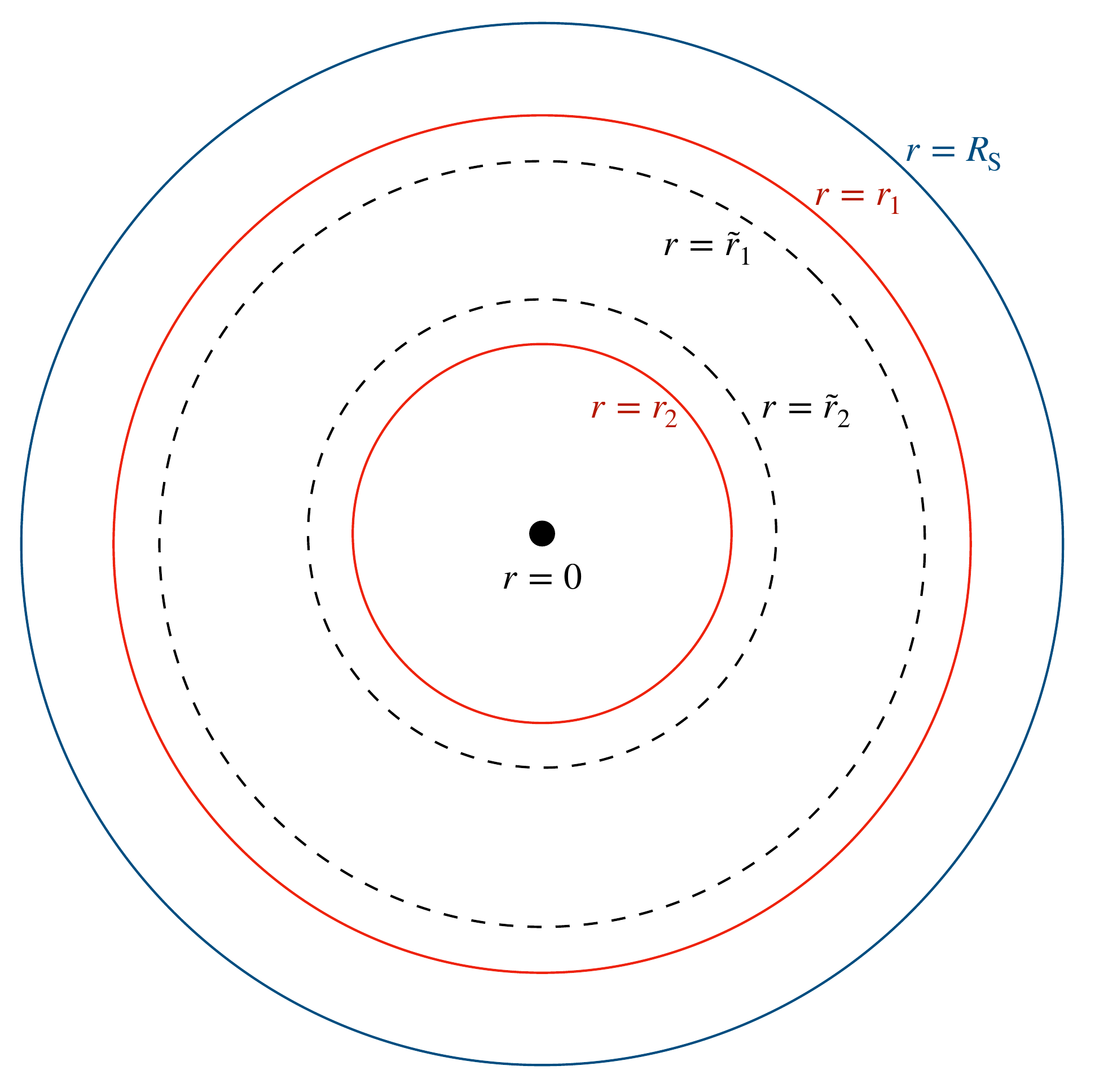}
\caption{Pictorial representation of the  Schwarzschild radius,  horizons located at $r=r_1$ and $r=r_2$, and null hypersurfaces  $\Sigma_1$ and $\Sigma_2$ in the case $k_1 >0$. }
\label{Fig-k1-positive}
\end{figure}

Let there be given in the spacetime geometry \eqref{Schwarzschild_metric_quantum-standard} an hypersurface, say $\Sigma$,  having constant radius and normal vector $n^\mu$. $\Sigma$ becomes null when 
\begin{align}
g_{\mu \nu} n^\mu n^\nu =g^{rr}=0,
\label{n-mu-n-mu}
\end{align}
where, owing to Eq. \eqref{A-of-r-quantum} and  up to $\OO\left(r^{-4}\right)$ terms,
\begin{align} \label{g-rr-inv-quatum}
g^{rr}=  1-\frac{\RS}{r} + \frac{3}{2}k_1 \frac{\RS \LP^2}{r^3}. 
\end{align}
In the standard Schwarzschild geometry, where $k_1 =0$, the condition $g^{rr}=0$ is equivalent to $g_{tt}=0$. Therefore, the (event) horizon located at $r=\RS$ is automatically a null hypersurface, i.e., the place where nothing, not even light, can escape from. This circumstance is no longer true in  the Schwarzschild geometry \eqref{Schwarzschild_metric_quantum-standard}, since, as pointed out before, $g^{rr}$ is different from  $-g_{tt}$. Therefore, a null hypersurface $\Sigma$ does not coincide with any of the metric horizons analyzed before. In fact, the relation  \eqref{n-mu-n-mu} leads to the new algebraic third-order equation $r^3 - r^2 \RS + \tfrac{3}{2} k_1 \RS \LP^2=0$, which admits two real positive roots when $k_1>0$ and $M > \sqrt{3} M^\star$, and one real positive solution whenever $k_1$ is negative.  In the first scenario, upon employing Eq. \eqref{Rs-bigger-than-lp}, the radii of the null hypersurfaces $\Sigma_1$ and $\Sigma_2$ are found to be
\begin{align}
\tilde{r}_1 &=  R_{\rm S} \left[1-\frac{3 k_1}{2} \frac{\LP^2}{\RS^2} + \OO\left(\ell_{\rm P}^4/R_{\rm S}^4\right)\right],
\label{solution-r-tilde-1}  
 \\
\tilde{r}_2 &= \LP \sqrt{\frac{3k_1}{2}} + \OO\left(\LP^2/\RS\right),
\label{solution-r-tilde-2}
\end{align}
respectively. In our hypotheses, we have
\begin{align}
\tilde{r}_2 \ll \tilde{r}_1,    
\end{align}
along with 
\begin{align}
\tilde{r}_1 &< r_1 < \RS,
\label{inequality-r1-tilde-r1}
\\
\tilde{r}_2 & > r_2.
\end{align}
The positions of the metric horizons \eqref{solution-r1} and \eqref{solution-r2} and the null hypersurfaces $\Sigma_1$ and $\Sigma_2$  are drawn in   Fig. \ref{Fig-k1-positive}.

\begin{figure}[bht!]
\centering\includegraphics[scale=0.35]{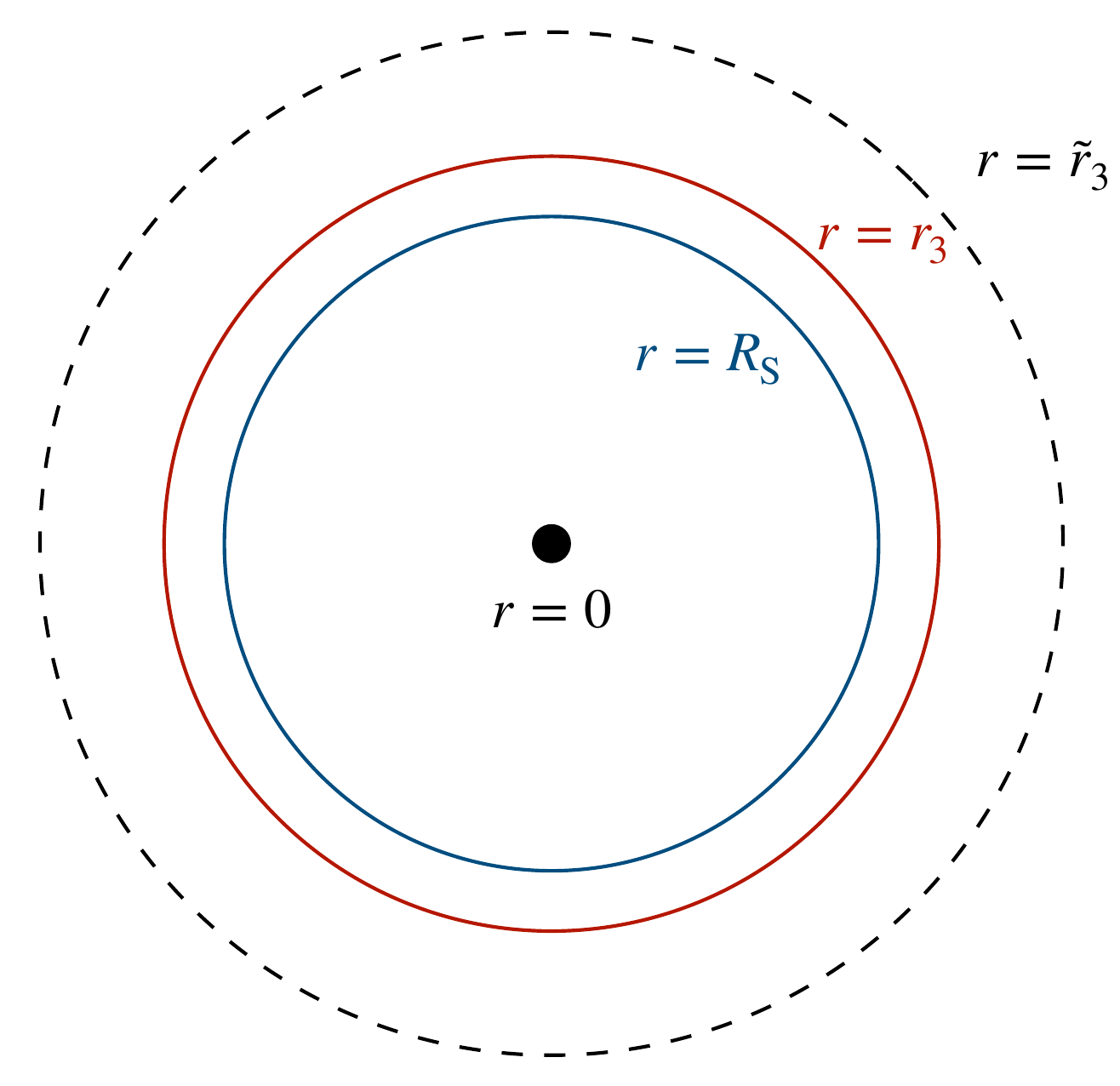}
\caption{Pictorial representation of  the null hypersurface  $\Sigma_3$, horizon located at $r=r_3$,  and  Schwarzschild radius in the case $k_1 <0$. }
\label{Fig-k1-negative}
\end{figure}

When $k_1 <0$, the null hypersurface $\Sigma_3$ is characterized by the radius
\begin{align}
\tilde{r}_3 &= R_{\rm S} \left[ 1-\frac{3k_1}{2} \frac{\LP^2}{\RS^2} +\OO\left(\ell_{\rm P}^4/R_{\rm S}^4\right) \right],
\label{solution-r-tilde-3}
\end{align}
and satisfies (cf. Eq. \eqref{solution-r3})
\begin{align}
\tilde{r}_3 >  r_3  > \RS.
\label{inequality-tilde-r3-r3}    
\end{align}
The main features of the framework having negative $k_1$ are shown in Fig. \ref{Fig-k1-negative}. 

The study of the sign of $g^{rr}$ reveals that \cite{Battista2023}
\begin{subequations}
\label{sign-grr-relations-1-2-3}
\begin{align}
&  r < \tilde{r}_2: \;\, \; \quad \quad \qquad g^{rr}>0, \label{sign-grr-1} \\
& \tilde{r}_2 < r < \tilde{r}_1: \qquad \;  \, g^{rr} < 0, \\
&  r >\tilde{r}_1: \qquad \qquad \, \, \; g^{rr} > 0, 
\end{align}
\end{subequations}
which entails that  all hypersurfaces  $\Sigma$ whose radius is smaller than $\tilde{r}_2$ or larger than $\tilde{r}_1$ are timelike and hence can be crossed by a particle either inwards or outwards \cite{Ferrari2020}; moreover, the hypersurfaces $\Sigma$ having a  radius lying in the interval $ \tilde{r}_2 <r<\tilde{r}_1$ are spacelike, i.e.,   they can be crossed by a particle in one direction only \cite{Ferrari2020}. On the other hand, in the case $k_1 <0$, we have the following situation:
\begin{subequations}
\label{sign-grr-relations-3-4}
\begin{align}
&  r < \tilde{r}_3: \;\; \qquad \quad \;\;\, g^{rr}<0,  \label{sign-grr-4}\\
&  r >\tilde{r}_3: \qquad \qquad \, \, g^{rr} > 0, \label{sign-grr-5} 
\end{align}
\end{subequations}
which means that the hypersurfaces $\Sigma$ are spacelike if $r <\tilde{r}_3$, while  they are timelike otherwise.

\subsection{Change of signature}\label{Sec:change-signature}

Among the peculiarities of the quantum Schwarzschild   metric \eqref{Schwarzschild_metric_quantum-standard}, we also remark the fact that its signature is allowed to change  within some Planck-scale domains. 

In the scenario having positive $k_1$,  $g_{tt}$ is negative if $r < r_2$ or $r > r_1$, while it attains  positive values otherwise \cite{Battista2023}; thus,   bearing in mind Eq. \eqref{sign-grr-relations-1-2-3},  we see that $g_{\mu \nu}$ assumes the Euclidean (i.e., Riemannian) signature $(++++)$ when $r_2 < r < \tilde{r}_2$ and $\tilde{r}_1 < r < r_1$ (see Fig. \ref{Fig-k1-positive}). On the other hand, if $k_1$ is negative, the temporal component $g_{tt}$ is positive (resp. negative) if $r<r_3$ (resp. $r>r_3$) \cite{Battista2023}, which, owing to Eq. \eqref{sign-grr-relations-3-4}, implies that  the metric admits the ultrahyperbolic signature $(--++)$ for $r_3 < r < \tilde{r}_3$ (see Fig. \ref{Fig-k1-negative}).

This behaviour of the metric can be ascribed to its quantum nature, which becomes more significant in the aforementioned Planck-extent regions. This point is also suggested by a comparison   with  some existing and well-studied quantum models. In fact, the regime with $k_1 >0$ is reminiscent of the change-of-signature idea featured in quantum cosmology, where the spacetime metric experiences a signature transition from  Euclidean to Lorentzian at the beginning of the cosmological evolution (see e.g. Refs. \cite{Gibbons1990,Dereli1993}); a similar  pattern is  also exploited in other fields, such as  loop quantum gravity  \cite{Bojowald2020}, supergravity and string theory \cite{Barrett1993,Mars2000,Stern2018}.  The framework with negative $k_1$ shares some similarities with a class of theories having two time dimensions, known as  two-time or 2T physics, which finds application in classical and quantum physics including field theories, gravity, and cosmology (see e.g. Refs. \cite{Bars1998,Bars2008a,Bars2010,Kamenshchik2023} and references therein).

\subsection{Effective energy-momentum tensor}\label{Sec:Effective-EMT}

The quantum Schwarzschild metric \eqref{Schwarzschild_metric_quantum-standard} can be interpreted as a non-vacuum solution of the Einstein field equations. In this context,  quantum contributions can be associated  with the presence of an  effective stress-energy tensor that permits gaining insights into the physical aspects of the underlying matter fields.   This approach  enables in general a deeper understanding of the interplay between quantum  effects and classical geometry, shedding light on the  properties of spacetimes involving  one-loop quantum modifications.

In order to perform our investigation,  we extend the metric given in Eq.~\eqref{Schwarzschild_metric_quantum-standard} as follows:
\begin{subequations}
\label{Schwarzschild_metric_quantum-highorder}
\begin{align}
{\rm d}s^2=\tilde{g}_{\mu \nu}{\rm d}x^\mu{\rm d}x^\nu= - \tilde{B}(r)\left(c^2{\rm d}t^2\right) + \tilde{A}(r){\rm d}r^2 + r^2 {\rm d}\Omega^2,
\end{align}
\text{with}  
\begin{align} 
\tilde{B}(r) &= 1-\frac{\RS}{r} + \frac{k_1}{2} \frac{\RS \LP^2 }{ r^3} + \frac{f_{\rm b}k_1 \RS ^2 \LP^2}{r^4}+\OO\left(r^{-5}\right),
\label{B-of-r-quantum-highorder}
\\
\tilde{A}(r) &= \left(1-\frac{\RS}{r} \right)^{-1} - \frac{3 k_1}{2} \frac{\RS \LP^2}{ r^3}+ \frac{f_{\rm a}k_1\RS ^2 \LP^2}{r^4}+\OO\left(r^{-5}\right),
\label{A-of-r-quantum-highorder}
\end{align}
\end{subequations}
where the explicit value of the constants $f_{\rm a}$ and $f_{\rm  b}$   entering the (leading) two-loop quantum corrections will not impact our examination. Starting from Eq.  \eqref{Schwarzschild_metric_quantum-highorder}, one could derive the corresponding Einstein tensor $E_{\mu \nu}$, defined as usual as
\begin{align}
    E_{\mu \nu} =R_{\mu \nu}-\frac{1}{2}g_{\mu \nu} R\,.
\end{align}
Considering the leading-order factors in $\LP ^2$, the $t-t$ and $r-r$ components of $E_{\mu \nu}$ read as, respectively, 
\begin{subequations}
\begin{align}
\tensor{E}{_t _t}=&\frac{3 k_1 \RS \LP^2}{r^5}-\frac{3 k_1 \RS ^2   \LP^2 \left(f_{\rm a}+4 \right)}{r^6}+\frac{k_1 \RS^3 \LP^2 \left(11 f_{\rm a}+15 \right)}{r^7}\notag \\
&-\frac{k_1 \RS^4  \LP^2 \left(13 f_{\rm a}+6 \right)}{r^8}+ \frac{5 f_{\rm a} k_1 \RS^5 \LP^2 }{r^9}\,,
\label{eq:Einstein tensor_ttfull} \\
\tensor{E}{_r _r}=&-\frac{2 k_1 \RS^2 \LP^2 \left(2 f_{\rm b}+1\right)}{r^4 (r-\RS)^2}+\frac{3 k_1 \RS^3 \LP^2 \left(2 f_{\rm b} +1\right)}{2 r^5 (r-\RS)^2}-\frac{f_{\rm a} k_1 \RS^2\LP ^2}{r^6}\,. 
\label{eq:Einstein tensor_rrfull}
\end{align}
\end{subequations}
From the above formulas, it is evident  that if we neglect two-loop  $\OO\left(r^{-4}\right)$  contributions 
in Eq.~\eqref{Schwarzschild_metric_quantum-highorder} (as we did in Eq.~\eqref{Schwarzschild_metric_quantum-standard}),  only the first factor on the right-hand side of Eq.~\eqref{eq:Einstein tensor_ttfull} should  be retained, while all terms occurring in  the expression \eqref{eq:Einstein tensor_rrfull} of $\tensor{E}{_r _r}$ should be discarded. 
By applying the same procedure also to $\tensor{E}{_\theta _\theta}$ and $\tensor{E}{_\phi _\phi}$ while keeping again the one-loop   accuracy in the metric \eqref{Schwarzschild_metric_quantum-highorder},  we find that these vanish. Therefore, the final form of $E_{\mu \nu}$ turns out to be
\begin{subequations}\label{eq:Einstein tensor}
\begin{align}
    \tensor{E}{_t _t}&=\frac{3 k_1 \RS \LP^2}{r^5}\,,\\
    \tensor{E}{_r _r}&=\tensor{E}{_\theta _\theta}=\tensor{E}{_\phi _\phi}=0\,,   
\end{align}
\end{subequations}
which represents  the correct Einstein tensor if one starts from the one-loop corrected geometry~\eqref{Schwarzschild_metric_quantum-standard}. Assuming that the Einstein gravitational field equations still hold in this situation, namely 
\begin{align}
    E_{\mu \nu} =\frac{8\pi G}{c^4} T_{\mu \nu}\,,
\end{align}
then from Eq.~\eqref{eq:Einstein tensor}  we arrive at the corresponding effective energy-momentum tensor:
\begin{align}\label{EMT_eff}
   \tensor{T} {_\mu _\nu}&=\frac{3 k_1  c^4 \RS \LP ^2}{8\pi G r^5}\,{\rm diag}\left(1,\,0,\,0,\,0 \right).
\end{align}
As a self-consistency check, one can verify that $\tensor{T} {_\mu _\nu}$ is conserved up to higher-order contributions. 

The knowledge of the effective energy-momentum tensor allows us to evaluate an important physical quantity like the effective energy density
\begin{align}
    \rho=-\tensor{T} {_t ^t}=\frac{3 k_1 \LP ^2 c^4 \RS}{8\pi G r^5}\,,
\end{align}
which we have written in such a  way as to be  identical to the energy density measured by a static observer in her/his local ``proper reference frame''  \cite{Morris-Thorne1988}. It is thus clear that $\rho$ is always negative if $k_1 <0$, a condition which points out that  the weak energy condition fails to hold \cite{Wald}. Moreover, we can also determine if the NEC is respected.  For this reason, let us introduce the  null vector 
\begin{align}
k^\mu =\left( 1/\tilde{B}, \,\pm1/\sqrt{\tilde{A}\tilde{B}},\,0,\,0\right)\,,
\label{radial-null-vector}
\end{align}
which points in the outgoing or ingoing radial direction, depending on whether the plus or minus sign is chosen, respectively.  Owing to Eqs. \eqref{EMT_eff} and \eqref{radial-null-vector}, we  obtain 
\begin{align}\label{eq:nec1}
T_{\mu \nu} k^{\nu}  k^{\nu}=\frac{3 k_1 \LP ^2 c^4 \RS}{8\pi G r^5 (1-\RS/r)^2}+\OO \left(r^{-6}\right)\,,
\end{align}
and hence we can conclude, modulo higher-order factors, that the NEC  is contravened for  negative $k_1$. 

The NEC violation is a phenomenon that typically characterizes physical systems involving quantum processes, and as such,  it is expected to occur also in the quantum Schwarzschild  geometry we are studying. Our investigation shows that, at this order, the NEC is satisfied for positive $k_1$, but the situation may change when next-to-leading-order terms are taken into account. For instance, if we just write $f_{\rm b} = \lambda f_{\rm a}$ (with $\lambda \in \mathbb{R}$) in Eq. \eqref{Schwarzschild_metric_quantum-highorder}, then the NEC could be breached also when  $k_1$ is positive for $r<[14 +4(1+\lambda) f_{\rm a}]\RS/3$. Therefore, we can say that the calculations carried out in this section represent a direct proof of the NEC infringement in the scenario having negative $k_1$, while an indirect evidence that NEC is broken when $k_1>0$ is provided by the existence of the Penrose-like energy-extraction mechanism discussed in Ref. \cite{Battista2023}. We will return to this point in Sec. \ref{Sec:radial-geodesics}. 

A final remark is now in order. The analysis of Ref. \cite{Bohr2003} has demonstrated how the Schwarzschild metric can be derived at one-loop level, starting from the energy-momentum tensor of a source that contains the contributions of both the classical gravitational field and its quantum fluctuations. Alternatively, in our approach, the metric \eqref{Schwarzschild_metric_quantum-highorder} consists of the exact ordinary classical piece plus small quantum corrections. As a result, the effective energy-momentum tensor \eqref{EMT_eff} (or equivalently the Einstein tensor \eqref{eq:Einstein tensor}) is a fully quantum object, where, differently from Ref. \cite{Bohr2003}, no classical factors are present.

\subsection{Thermodynamic aspects}\label{Sec:thermod-aspects}

Having outlined the general features of the quantum geometry \eqref{Schwarzschild_metric_quantum-standard}, we now provide a first examination of its thermodynamic aspects. Specifically, we compute the black hole entropy using two approaches: one relying on the Hawking temperature (see Sec. \ref{Sec:Hawking-temperature}) and the other on the Wald formula (see Sec. \ref{Sec:Wald-formula}).

\subsubsection{Hawking temperature} \label{Sec:Hawking-temperature}

The Hawking temperature $T_{\rm H}$ is a fundamental concept in black hole thermodynamics that reflects the interaction between gravity and quantum field theory in curved spacetimes \cite{Wald-LRR}. It can be formally expressed as \cite{Wald}
\begin{align}
T_{\rm H} = \frac{\hbar}{c k_{\rm B}}\frac{\kappa}{2\pi},
\end{align}
$\kappa$ being the surface gravity. In a spherically symmetric spacetime, we can write
\begin{align}
\kappa^2 = c^4\left[-\frac{1}{2 } \left(\nabla_\mu \xi_\nu\right) \left(\nabla^\mu \xi^\nu\right) \right]_{r=r_H}=c^4\left[ -\frac{1}{4 } \frac{1}{g_{rr} g_{tt}}   \left(g_{tt}^\prime\right)^2 \right]_{r=r_H},
\end{align}
where $\xi^\mu = \delta^{\mu 0}$ denotes the static Killing vector field, the prime  the derivative with respect to $r$, and $r_H$ the horizon radius, which in our setup is located at $r_1$ when $k_1>0$ and at $r_3$ if $k_1<0$. It follows from Eqs. \eqref{B-of-r-quantum} and \eqref{A-of-r-quantum}, that 
\begin{align}
 -\frac{1}{4 } \frac{1}{g_{rr} g_{tt}}   \left(g_{tt}^\prime\right)^2  = \frac{\RS^4}{4} \left[\frac{1}{\RS^2 }\frac{1}{r^4}-3 k_1 \frac{\LP^2}{\RS^2} \frac{1}{r^6} + \OO \left(r^{-7}\right)\right],   
\end{align}
which, in view of Eqs. \eqref{solution-r1} and \eqref{solution-r3},  yields
\begin{align}
T_{\rm H} = \frac{\hbar c}{ k_{\rm B}}\frac{1}{4\pi \RS} \left[1-\frac{k_1}{2 } \frac{\LP^2}{\RS^2} + \OO\left(\LP^4/\RS^4\right)\right],
\end{align}
where we have taken into account the condition \eqref{Rs-bigger-than-lp}. 
If we now use the first law of thermodynamics $c^2 \dd M = T_{\rm H} \dd \mathcal{S}$, we then obtain that the entropy is given by, modulo  higher-order corrections,
\begin{align}
\mathcal{S} = c^2 \int \frac{\dd M}{T_{\rm H}(M)}= \frac{\pi k_{\rm B} c^3}{\hbar G} \RS^2 \left[1 + k_1 \frac{\LP^2}{\RS^2} \log \left(\frac{\RS}{\LP}\right)\right].
\label{entropy-approach1}
\end{align}
This equation suggests that $\LP$, or equivalently $\MP$, can be seen as a natural cutoff for the model, consistently with the EFT paradigm (a similar consideration will be taken into account   in  Sec. \ref{Sec:quartic-analysis-quantum}). 

Let us notice that, remarkably,  a correction to the entropy involving a logarithmic  term is present also in the general formulas worked out in Ref. \cite{DelPiano2023}, and it is expected to be a general feature of  EFT black hole models (see e.g. Refs. \cite{Fursaev1994,Cai2009,El-Menoufi2015,Calmet2021b,Xiao2021}).

\subsubsection{Wald entropy formula}\label{Sec:Wald-formula}

We can now calculate the black hole entropy $\mathcal{S}$  using the renowned Wald  formula \cite{Wald1993,Iyer1994,Wald-LRR}. In this framework,  $\mathcal{S}$ is elegantly determined through an integral expression involving the Noether charge associated with the diffeomorphism invariance of a general gravity model.  Specifically, for a diffeomorphism-invariant action 
\begin{align}
\mathcal{I} = \int  \dd^4 x \sqrt{-g} \, \mathcal{L}\left(g_{\alpha \beta},R_{\mu \nu\rho\sigma}, \nabla_{\delta}R_{\mu \nu\rho\sigma}, \Psi, \nabla_\delta \Psi\right),
\end{align}
with $\Psi$ the collection of matter fields in the theory and $R_{\mu \nu \rho \sigma}$ the Riemann tensor, the entropy is  given by
\begin{align}\label{Wald-entropy-formula}
 \mathcal{S} = - 2 \pi \frac{k_{\rm B}}{\hbar c} \int \limits_{r=r_H} \dd \Sigma \, \epsilon_{\mu \nu} \epsilon_{\rho \sigma} \frac{\delta \mathcal{L}}{\delta R_{\mu \nu \rho \sigma}}    , 
\end{align}
where, for a spherically symmetric and static spacetime, $\dd \Sigma = r^2 \sin \theta \,  \dd \theta \dd \phi$, and $\epsilon_{\mu \nu}$ is an antisymmetric tensor, normalized to $\epsilon_{\mu \nu} \epsilon^{\mu \nu}=-2$, whose nonvanishing component is $\epsilon_{tr} = \sqrt{AB}$ (cf. Eq. \eqref{Schwarzschild_metric_quantum-standard})\footnote{Strictly speaking,  Wald formula applies to  bifurcate Killing horizons of  stationary black holes and $\epsilon_{\mu \nu}$ is the binormal vector to the bifurcation surface.}. 

The action suitable for the EFT  treatment of general relativity includes both local  higher-order curvature corrections to the Einstein-Hilbert Lagrangian and nonlocal contributions, and thus reads as \cite{Donoghue1994,Calmet2019}
\begin{align}
\mathcal{I}_{\rm EFT} &= \int  \dd^4 x \sqrt{-g} \, \left[ \chi R + c_1 R^2 + c_2 R_{\alpha \beta} R^{\alpha \beta} + c_3 R_{\alpha \beta \varepsilon \omega} R^{\alpha \beta \varepsilon \omega} + \mathcal{L}_{\rm m} \right.
\nonumber \\
&\left. -\vartheta R_{\alpha \beta \varepsilon \omega} \log \left(\Box/\varrho^2\right) R^{\alpha \beta \varepsilon \omega} + \dots \right],
\label{EFT-action-1}
\end{align}
where $\chi= c^4 \left(16 \pi G\right)^{-1}$, $c_{1,2,3} $ and $ \vartheta$ are constants, $\Box:= g^{\mu \nu} \nabla_\mu \nabla_\nu$ denotes the d'Alembertian operator, $\mathcal{L}_{\rm m}$ the matter Lagrangian, $\varrho^{-1}$ a reference length scale, and the dots indicate higher-order modifications. The nonlocal factor involving  $\log \left(\Box/\varrho^2\right)$ is responsible for the logarithmic term occurring in Eq. \eqref{entropy-approach1}. In fact, applying the Wald formula \eqref{Wald-entropy-formula}  to the EFT action \eqref{EFT-action-1} we find that the entropy of the quantum Schwarzschild black hole takes the form
\begin{align}
\mathcal{S} & = \frac{8 \pi^2 k_{\rm B}}{\hbar c} r_H^2 \left\{  2 \chi + 4 c_1 \left(R\right)_{r_H} - 2 c_2 \left(\tensor{R}{_\mu^\rho} \epsilon^{\mu\sigma} \epsilon_{\rho \sigma}\right)_{r_H} \right.
\nonumber \\
& \left.    - 2 \left(\epsilon_{\mu \nu} \epsilon_{\rho \sigma} R^{\mu \nu \rho \sigma}\right)_{r_H} \left[c_3 + 2 \vartheta \left(\gamma_{\rm E}-1 \right) + \vartheta \log \left(r_H^2 \varrho^2\right)\right] \right\},
\label{entropy-Wald-1}
\end{align}
where  the functional derivatives  have been evaluated by  resorting to the relation (see Appendix L in Ref. \cite{Teimouri2018} for a glossary of functional differentiation formulas)
\begin{align}
    \frac{\delta R_{\alpha \beta \gamma \xi}}{\delta R_{\mu \nu \rho \sigma}} = \frac{1}{4} \left(\delta^\mu_\alpha \delta^\nu_\beta-\delta^\nu_\alpha \delta^\mu_\beta\right) \left(\delta^\rho_\gamma \delta^\sigma_\xi- \delta^\sigma_\gamma \delta^\rho_\xi \right), 
\end{align}
and the last term follows from the identity  
\begin{align}
  \log \left(\Box/\varrho^2\right) R^{\alpha \beta \varepsilon \omega}=  -R^{\alpha \beta \varepsilon \omega} \left[\log \left(r^2 \varrho^2\right) + 2 \left(\gamma_{\rm E} -1\right)\right],
  \label{identity-log-box}
\end{align}
$\gamma_{\rm E}$ being the Euler-Mascheroni constant (see e.g. Refs. \cite{Calmet2017,Calmet2019,Calmet2021b,Xiao2021} for further details on the derivation of Eq. \eqref{identity-log-box}). Then, owing to the relations (cf. Eqs. \eqref{solution-r1} and \eqref{solution-r3})
\begin{subequations}
\begin{align}
\left(r^2 R \right)_{r_H}&= \left[\frac{3 k_1 \LP^2 \RS}{r^3}+ \OO{\left( r^{-4}\right)}\right]_{r_H}=3 k_1 \frac{\LP^2}{\RS^2} + \OO(\LP^4/\RS^4), 
\\
 \left(r^2 \tensor{R}{_\mu^\rho} \epsilon^{\mu\sigma} \epsilon_{\rho \sigma}\right)_{r_H}&=\left[ \frac{19 k_1 \LP^2 \RS^2}{2 r^4}+\OO{\left(r^{-5}\right)}\right]_{r_H}=  \frac{19k_1\LP^2}{2\RS^2} + \OO(\LP^4/\RS^4), 
 \\
 \left(r^2 \epsilon_{\mu \nu} \epsilon_{\rho \sigma} R^{\mu \nu \rho \sigma}\right)_{r_H}&=\left[-\frac{4\RS}{r}+\frac{12k_1\RS \LP^2}{r^3}+\OO{\left(r^{-4}\right)}\right]_{r_H}=-4  +  \frac{10 k_1 \LP^2}{\RS^2}+ \OO(\LP^4/\RS^4), 
\end{align}
\end{subequations}
we obtain from Eq. \eqref{entropy-Wald-1}  
\begin{align}
\mathcal{S} &=  \frac{8 \pi^2 k_{\rm B}}{\hbar c} \RS^2  \Biggl\{ 2 \chi \left[1-\frac{k_1 \LP^2}{\RS^2}+\OO{\left(\LP^4/\RS^4\right)}\right] +\frac{k_1 \LP^2}{\RS^4} \left(12 c_1 - 19 c_2\right) + \Bigl[8/\RS^2 - 20 k_1 \LP^2/\RS^4 
\nonumber \\
&+ \OO{\left(\LP^4/\RS^6\right)}\Bigr]  \left[c_3 + 2 \vartheta \left(\gamma_{\rm E}-1\right) + \vartheta \log\left(r_H^2 \varrho^2\right)\right] + \OO{\left(\LP^4/\RS^6\right)} \Biggr\}.
\end{align}
Since  the Planck length is expected to be the characteristic scale of the ultimate theory of quantum gravity,  we can set $\varrho^2 \equiv \LP^{-2}$.  Discarding higher-order corrections, we thus obtain
\begin{align}
\mathcal{S} =  \frac{\pi k_{\rm B} c^3}{\hbar G} \RS^2  \left\{ 1+ \frac{4}{\chi \RS^2} \left[c_3 + 2 \vartheta \left(\gamma_{\rm E}-1\right) \right] + \frac{8\vartheta}{\chi} \frac{1}{\RS^2} \log\left(\frac{\RS}{\LP}\right)\right\},
\label{entropy-Wald-2}
\end{align}
which implies that the sought-after expression \eqref{entropy-approach1} for   the entropy is recovered if 
\begin{align}
c_3 &= -2 \vartheta (\gamma_{\rm E}-1),
\nonumber \\
\vartheta &= \frac{k_1 \LP^2}{8} \chi,
\end{align}
while $c_1$ and $c_2$ can assume any value since  they are related to terms yielding, at this level, negligible corrections. We notice that, remarkably, our final result aligns with the outcomes of e.g. Ref. \cite{Calmet2021b}.

A final remark deserves mention. The approach followed in this section does not represent the only way  to derive the entropy formula  \eqref{entropy-approach1}. For instance, the class of local operators of the type $R_{\alpha \beta } \log \left(\Box/\varrho^2\right) R^{\alpha \beta }$ or $R \log \left(\Box/\varrho^2\right) R$ can also enter $\mathcal{I}_{\rm EFT}$ and  lead to the logarithmic factor  in $\mathcal{S}$. Nonetheless, the calculations presented here serve as a valuable double-check of the effectiveness of this  contribution.

\section{Geodesic dynamics of massive and massless particles}\label{Sec:geodesic-motion-massive-massless}

\begin{figure}
    \centering
    \begin{subfigure}[t]{0.49\textwidth}
   \includegraphics[width=7.0cm]{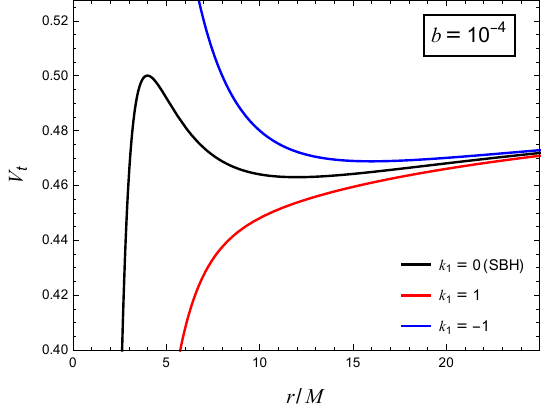}
  \end{subfigure} 
    \begin{subfigure}[t]{0.49\textwidth}
      \includegraphics[width=7.0cm]{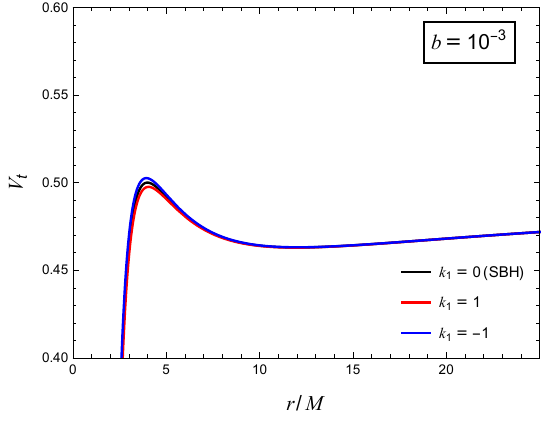}
  \end{subfigure}  
    \caption{Effective potential $V_{\rm t}(r)$  for massive particles with $G=1$, $\LP/\RS=10^{-4}$, $L=4M$, and different values of $k_1$ (see Eq. \eqref{effective-potential-general} with $\alpha=1$).  The black lines refer to the classical Schwarzschild black hole,  abbreviated as  \qm{SBH}, where $V_{\rm t}$ admits  a maximum at $r=4M$ and a minimum at $r=12M$. In the left panel, $V_{\rm t}$ has no extremum when $k_1 = 1$, while for $k_1 =-1$ it reaches  the minimum at $r = 16M$. In the right panel, $V_{\rm t}$ exhibits a maximum at $r=4.06M$ and a minium at $r=11.94M$ for $k_1=1$, while for $k_1=-1$ these extrema are achieved   at $r=3.94M$ and  $r=12.06M$, respectively. }
    \label{fig:Vt_eff}
\end{figure}

\begin{figure}
    \centering
    \begin{subfigure}[t]{0.49\textwidth}
   \includegraphics[width=7.0cm]{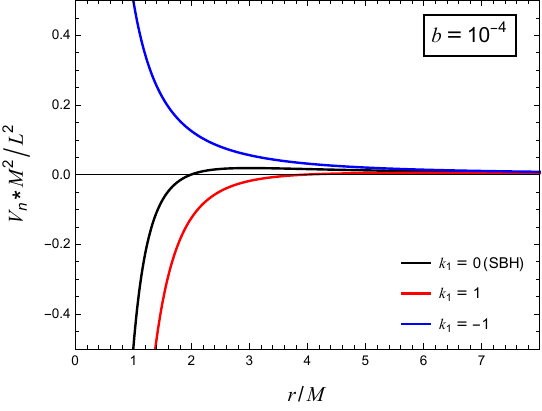}
  \end{subfigure} 
    \begin{subfigure}[t]{0.49\textwidth}
      \includegraphics[width=7.0cm]{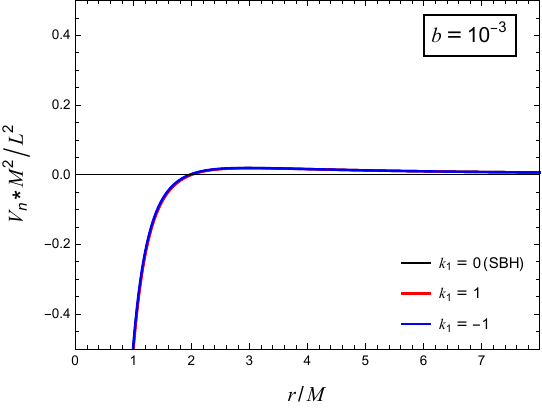}
  \end{subfigure}  
\caption{Effective potential $V_{\rm n}(r)$ for massless particles with $G=1$, $\LP/\RS=10^{-4}$, and  $k_1=0,\pm 1$ (see Eq. \eqref{effective-potential-general} with $\alpha=0$). For the classical Schwarzschild black hole (SBH),   $V_{\rm n}$ attains its maximum value at the $b$-independent point $r = 3M$.  In the left panel,  $V_{\rm n} $ admits no maximum when $k_1 = -1$, while for $k_1 = 1$ it is reached at $r = 6M$, which  does not correspond to  the photon sphere radius due to the energy dependence in the effective potential (a more detailed discussion will be provided in Sec.~\ref{Sec:null-geodesics}). In the right panel, the quantum and classical effective potentials  almost overlap, with   the maximum of $V_{\rm n} $ occurring at  $r = 3.03M$ for $k_1 = 1$ and   $r = 2.97M$ for $k_1 = -1$.  }
\label{fig:Vn_eff}
\end{figure}

The geodesic path of a freely falling particle moving in the equatorial plane of the  quantum  geometry \eqref{Schwarzschild_metric_quantum-standard} can described starting from  the equation (hereafter, we set $c=1$)  \cite{Battista2023}
\begin{align}
 \frac{1}{2} \dot{r}^2 + \frac{1}{2} \frac{1}{A(r)} \left(\frac{L^2}{r^2}+ \alpha \right)   = \frac{1}{2} \frac{\mathscr{E}^2}{A(r)B(r)},
 \label{geod-timelike-and-null-1}
\end{align}
where the dot stands for differentiation with respect to the affine parameter, and $\alpha=0,1$. The case $\alpha=1$ refers to timelike geodesics, where the constants of motion $\mathscr{E}$ and $L$ represent the energy and angular momentum per unit rest mass of a massive particle, respectively; $\alpha=0$ pertains to null geodesics, for which $\mathscr{E}$ and $L$ denote the total energy and angular momentum of a photon, respectively. Once  the terms beyond one loop (i.e., those of the order $\OR{-4}$) are neglected, Eq. \eqref{geod-timelike-and-null-1} yields
\begin{align}
\frac{1}{2} \dot{r}^2 + V_{\rm eff}(r) = \frac{1}{2} \mathscr{E}^2,    
\label{geod-timelike-one-loop}
\end{align}
where the quantum corrected effective potential is given by 
\begin{align}
V_{\rm eff}(r) = \frac{1}{2} \left(1-\frac{\RS}{r}\right) \left(\frac{L^2}{r^2}+ \alpha \right) + \frac{1}{2} k_1 \left(\frac{3 \alpha }{2} - \mathscr{E}^2 \right) \frac{\RS \LP^2}{r^3}.     
\label{effective-potential-general}
\end{align}
The behavior of $V_{\rm eff}(r)$  for  both massive and massless particles is displayed  in Figs.   \ref{fig:Vt_eff} and \ref{fig:Vn_eff}, respectively. Here, we have defined the impact parameter as $b:= L/\mathscr{E}$, with   its physical significance to be discussed in Sec.~\ref{Sec:Shadows-rings}. For the chosen parameters, we see that the quantum effective potential exhibits differences compared to the classical Schwarzschild black hole case that become more pronounced as $b$ decreases. This aligns with our expectations, since smaller values of $b$ correspond to particles getting closer to the black hole and experiencing the strong-field regime,  where quantum contributions gain  importance. 

At this point, some comments are in order. It is clear that Eq. \eqref{geod-timelike-and-null-1} receives some divergent contributions which can be interpreted as an infinite-depth barrier as soon as either $A$ or $B$, or both, are zero (the first situation occurs at the metric horizons $r_i$, with $i=1,2,3$, where $B(r_i)=0$ and $  A(r_i) \neq 0$). This feature cannot be inferred from the expressions \eqref{geod-timelike-one-loop} and  \eqref{effective-potential-general}, as  their derivation relies  only on one-loop quantities stemming from $A^{-1}$ and $(AB)^{-1}$. If we had an \emph{exact} formula for both $A$ and $B$, then   
the analysis of the geodesic motion would have required the use of  Eq.  \eqref{geod-timelike-and-null-1}, rather than Eqs. \eqref{geod-timelike-one-loop} and \eqref{effective-potential-general}. However,  an exact form of $A$ and $B$ can only be obtained via a full-loop calculation, which makes sense in a fully quantum regime. In this framework, the geodesic equation does not give a reliable 
description of the dynamics of a freely falling particle\footnote{See e.g. Refs. \cite{Bohr2015,Bohr2016,Bai2017,Chi2019}, where it is suggested that at quantum level some classical formulations of the equivalence principle seem to be violated.} and hence one should resort to quantum-mechanics tools, which we expect  not to be plagued by the presence of an infinite-depth barrier. 
However, since in this paper we only deal with the first low-energy quantum modifications of the classical Schwarzschild solution, a first approach where general-relativity techniques are employed for the study of the dynamics  is reasonable. 

We begin our investigation with timelike geodesics. In Sec. \ref{Sec:Timelike-geod}, we show that in the quantum realm the positions of stable and unstable circular orbits are described by a quartic algebraic equation, for which we propose an analytic  method to find its  roots; as we will see, the solutions are written in an approximated  form which is consistent with the EFT recipes and the ensuing constraint \eqref{Rs-bigger-than-lp}. After that, massive bodies trajectories are considered in Sec. \ref{Sec:trajectory-timelike}. Finally, we delve into null geodesics in Sec.  \ref{Sec:null-geodesics}, where we first exploit the aforementioned analytic technique to work out the quantum corrected photon sphere radii, and then examine the motion of photons. Our analysis unveils that the effect of the quantum contributions on the  orbits of both massive and massless particles  can give rise to intriguing phenomena.

\subsection{Timelike geodesics: stable and unstable circular orbits}\label{Sec:Timelike-geod}

A first preliminary study of stable and unstable circular orbits and, in particular, of the quantum corrections to the innermost stable circular orbit (ISCO) radius,  has been given in Ref. \cite{Battista2023}. In this section, we extend this former examination by providing an analytic method for obtaining in a  straightforward way the quantum corrected radii of such trajectories for both positive and negative $k_1$ scenarios. 

The analysis of timelike geodesics can be performed via Eq. \eqref{geod-timelike-one-loop}, where the effective potential should be evaluated by setting $\alpha=1$. In this way, from Eq. \eqref{effective-potential-general} we obtain the \qm{timelike potential} (see Fig. \ref{fig:Vt_eff})
\begin{align}
V_{\rm t}(r) = \frac{1}{2} \left(1-\frac{\RS}{r}\right) \left(\frac{L^2}{r^2}+1\right) + \frac{1}{2} k_1 \left(\frac{3}{2} - \mathscr{E}^2 \right) \frac{\RS \LP^2}{r^3},    
\label{effective-potential-timelike}
\end{align}
whose first-order and second-order derivatives with respect to the radial variable (denoted with a prime) read as 
\begin{subequations}
\begin{align}
V^\prime_{\rm t}(r) &= \frac{1}{r^4} \left[\frac{r^2 \RS}{2} + L^2 \left(\frac{3}{2} \RS-r \right) + \frac{9}{4} k_1 \RS \LP^2 \left(\frac{2}{3} \mathscr{E}^2 -1\right)\right],
\label{derivative-effective-potential-timelike}
\\
V_{\rm t}^{\prime \prime}(r) &=   \frac{1}{r^5} \left[3 L^2 (r-2 \RS)-\RS r^2 -9  k_1 \RS \LP^2 \left(\frac{2}{3} \mathscr{E}^2-1\right)\right],
\label{second-derivative-potential-timelike}
\end{align}
\end{subequations}
respectively. Let $E:= \mathscr{E}(\R)$ denote the value of the energy (per unit rest mass) of a particle following a circular orbit of radius $\R$, where $\R $ represents one of the extrema of $V_{\rm t}(r)$,  which thus satisfies $V^\prime_{\rm t}(\R)=0$. From the well-known relation $(1/2) E^2 = V_{\rm t} (\R)$ (cf. Eq. \eqref{geod-timelike-one-loop}),  we find modulo $\OO\left(\R^{-4} \right)$ contributions 
\begin{align}
E^2 = 1-\frac{\RS}{\R} + \frac{L^2}{\R^2}-\frac{L^2 \RS}{\R^3} + \frac{k_1}{2} \frac{\RS \LP^2}{\R^3}.
\label{E-squared-1}
\end{align}
We can further simplify this equation by exploiting the condition $V^\prime_{\rm t}(\R)=0$, which gives an expression for the ratio $L^2/\R^2$. In this way, by neglecting higher-order terms, we obtain from Eq.~\eqref{E-squared-1} the following formula pertaining to the quantum corrected  energy: 
\begin{align}
E=    \left[  \frac{2 \left(\R -\RS\right)^2}{\R \left(2\R -3\RS\right)}+\left(k_1 \RS \LP^2\right)  \frac{12 \left(\R -\RS\right)^3 + \R \left(6\RS -7 \R \right)\left(2\R -3\RS\right) }{2 \R^4 \left(2\R -3\RS\right)^2}\right]^{1/2}.
\label{energy-timelike}
\end{align}

By means of Eq. \eqref{E-squared-1},  the relation $V^\prime_{\rm t}(\R)=0$ gives   the fifth-order algebraic  equation 
\begin{align}
&\left(\RS \right) \R^5 -\left(2L^2\right) \R^4 + \left(3 \RS L^2 \right) \R^3 + \frac{3}{2} k_1 \RS \LP^2 \left[-\R^3 - \left(2 \RS\right) \R^2 +  \left(2 L^2\right) \R -2 L^2 \RS \right.
\nonumber \\
& \left. + k_1 \RS \LP^2 \right]=0.
\label{quintic-timelike}
\end{align}     
Quintic equations are characterized  by a  rich mathematical theory. The Abel-Ruffini theorem asserts that the solutions of  the equations of degree five or higher over the rationals cannot be given in terms of radicals, and hence more involved functions will occur in their resolution process. An example is furnished by the Birkeland algorithm, which expresses the roots of any quintic via generalized hypergeometric functions. Such procedure can be employed with notable  success once the quintic has been brought to so called Bring-Jerrard  form, i.e.,  it can be written as $X^5 + aX+d=0$.  This reduction can always be done by introducing a polynomial transformation known as  Tschirnhaus transformation, which is defined   in such a way as to  remove the factors of degree four, three, and two from the quintic. Another notable method to address fifth-order equations is  due to Hermite, who has proved  that a quintic in the Bring-Jerrard form can be solved by means of elliptic functions. Moreover, more recently a valuable  technique has been set forth in Ref. \cite{King1991}, where it is shown how the roots of the quintic  can be represented through  the Jacobi nome and the theta series. For further details regarding  fifth-degree equations and their applications in quantum settings we refer the reader to Refs. \cite{Battista2015a,Battista-book2017} and references therein. 

It is thus clear that the resolution of the quintic  \eqref{quintic-timelike} inevitably involves special functions. Therefore, attempting to decrease its degree  can provide substantial advantages. Remarkably, this reduction can be achieved within our model. Indeed, if we use Eq. \eqref{E-squared-1} to express $L^2$ in terms of $E^2$ and $\mathcal{R}$, we obtain
\begin{align}
L^2=\frac{\mathcal{R}^3 E^2-k_1 \RS \LP^2/2}{\mathcal{R}-\RS}-\mathcal{R}^2 \,, 
\label{L-2-and-E-timelike}
\end{align}
with which we can write the condition $V^\prime_{\rm t}(\R)=0$ as the following quartic equation:
\begin{align}
&\left(1-E^2\right)\R ^4 +\RS \left(\frac{3}{2}E^2-2\right) \R^3 + \RS ^2 \R^2 + \frac{1}{2} k_1 \RS \LP^2\left(3 E^2-\frac{7}{2}\right) \R 
\nonumber \\
&+\frac{3}{2}  k_1  \RS ^2 \LP^2 \left(1-E^2\right)=0. 
\label{quartic-timelike}
\end{align}

The physically meaningful solutions of this equation are the positive real-valued radii  giving 
positive values of $E$ and $L^2$ (see Eqs. \eqref{E-squared-1}, \eqref{energy-timelike}, and  \eqref{L-2-and-E-timelike}). Stable circular orbits (SCOs) having radius $\R$ are singled out by relation $V_{\rm t}^{\prime \prime}(\mathcal{R})\geq 0$ (cf. Eq. \eqref{second-derivative-potential-timelike}), where we will show that, similarly to the classical Schwarzschild framework, the equal sign is attained at the ISCO; on the other hand,  for unstable circular orbits (UCOs) the condition $V_{\rm t}^{\prime \prime}(\mathcal{R})< 0$ holds.   We first solve the quartic \eqref{quartic-timelike} in the classical case in Sec. \ref{Sec:quartic-analysis-classic}, which prepares the ground for the  investigation in the quantum realm, to be performed in Sec. \ref{Sec:quartic-analysis-quantum}.

\subsubsection{Analysis of the quartic equation in the classical regime}\label{Sec:quartic-analysis-classic}

In the classical regime $k_1=0$, the quartic \eqref{quartic-timelike}, which, as pointed out before,  governs the position of both SCOs and UCOs,  boils down to a second-order equation. In order to make contact with the notations to be introduced in Sec. \ref{Sec:quartic-analysis-quantum}, we write the ensuing roots in the following way:
\begin{subequations}
\label{bar-R-3-and-4}
\begin{align}
\bar{\mathcal{R}}_3 &=\frac{\left(3 \bar{E}^2-4+\bar{E}\sqrt{9 \bar{E}^2-8}\right) \RS} {4 \left(\bar{E}^2-1\right)},
\label{bar-R-3}
 \\
\bar{\mathcal{R}}_4 &=\frac{\left(3 \bar{E}^2-4-\bar{E}\sqrt{9 \bar{E}^2-8}\right) \RS} {4 \left(\bar{E}^2-1\right)},
\label{bar-R-4}
\end{align}
\end{subequations}
where $\bar{E}$ is the classical energy which can be read off from Eq. \eqref{energy-timelike} with $k_1=0$. It is simple to show  \cite{Wald} that, when $L^2 >12 M^2$, $\bar{\R}_3$ is the maximum while $\bar{\R}_4$ the minimum of the classical effective potential, which can be readily deduced from Eq. \eqref{effective-potential-timelike}. While  $\bar{\mathcal{R}}_4$    is defined only if $\bar{E}^2 \neq 1$,  $\bar{\mathcal{R}}_3$  admits a continuous extension at $\bar{E}=1$, where it attains the value 
\begin{align}
\bar{\R}^{(\bar{E}=1)}_{3}=:\bar{\R}_{+}=   2 \RS,    
\label{bar-R-plus}
\end{align}
which can be  easily obtained also from the quartic \eqref{quartic-timelike} in the limit $k_1=0$ and $E^2 \equiv \bar{E}^2  =1$.

\begin{figure}[bht!]
\centering\includegraphics[scale=0.70]{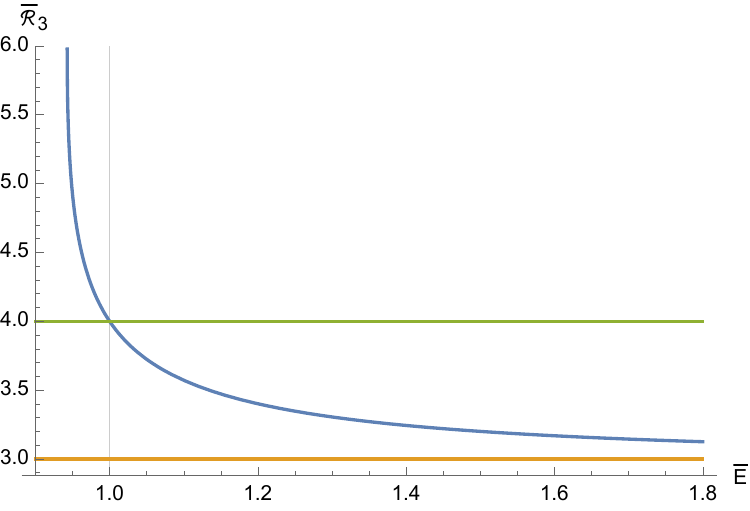}
\caption{The solution $\bar{\R}_{3}$ as a function of $\bar{E}$ with  $M=G=1$ (see Eq. \eqref{bar-R-3}). It is clear   that $\bar{\R}_{3}$ lies between $3M$ and $6M$ and that $\bar{E}^2$ becomes smaller than one as soon as $\bar{\R}_{3} >4M$. Moreover, $\bar{\R}_{3}$ admits a continuous extension at $\bar{E}=1$, as reported in Eq. \eqref{bar-R-plus}.}
\label{Fig:R-minus-E-classic}
\end{figure}

\begin{figure}[bht!]
\centering\includegraphics[scale=0.70]{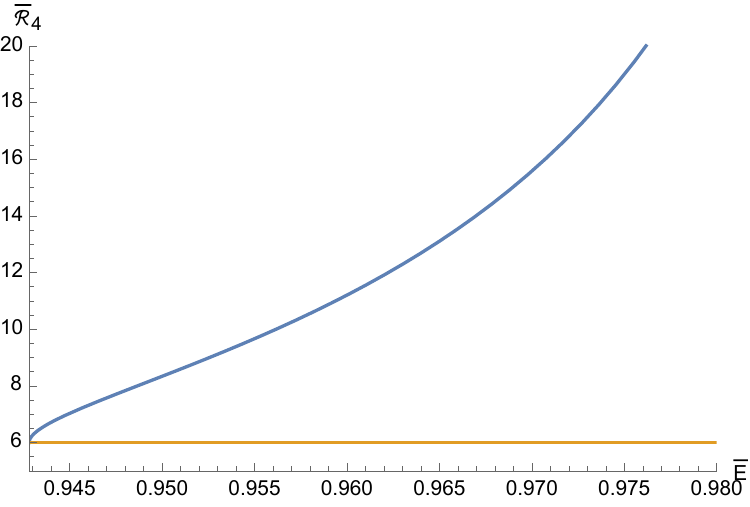}
\caption{The solution $\bar{\R}_{4}$ as a function of $\bar{E}$ with  $M=G=1$ (see Eq. \eqref{bar-R-4}). It is evident that  $\bar{\R}_{4} \geq 6M$  and $8/9 \leq \bar{E}^2 <1 $. }
\label{Fig:R-plus-E-classic}
\end{figure}

The plots of  $\bar{\R}_{3}$ and $\bar{\R}_{4}$ are displayed in Figs. \ref{Fig:R-minus-E-classic} and \ref{Fig:R-plus-E-classic}, respectively, 
where the following well-known classical results are evident \cite{Wald,Chandrasekhar1985}: 
\begin{subequations}
\begin{align}
&\frac{3}{2} \RS <\bar{\R}_{3} \leq  3 \RS, 
\label{classical-bound-1}
 \\
& \bar{\R}_{4} \geq 3 \RS,  
\label{classical-bound-2}
\\
& \frac{8}{9} \leq \bar{E}^2<1,  \quad {\rm if} \quad   \bar{\R} > 2 \RS, 
\label{classical-bound-3}
\\
&\bar{E}^2 \geq 1, \quad {\rm if} \quad \frac{3}{2} \RS < \bar{\R}_3 \leq 2 \RS,
\end{align}
\end{subequations}
where in Eq. \eqref{classical-bound-3} $\bar{\R}$ denotes either $\bar{\R}_3$ or $\bar{\R}_4$. The lower bound $\bar{E}^2  \geq 8/9$ guarantees that  $\bar{\R}_{3,4}$ are well-defined functions; in particular, $\bar{\R}_3$ attains its minimum value when $\bar{E} \gg1$, as
\begin{align}
\lim \limits_{\bar{E} \to +\infty} \bar{\R}_3 = \frac{3}{2} \RS;
\label{limit-classic-unstable}
\end{align}
moreover, SCOs with radius $ \bar{\R}_{4}$ have always $8/9 \leq \bar{E}^2<1$; lastly, when 
\begin{align}
\bar{E}^2=8/9=:\bar{E}_{\rm ISCO}^2,  
\label{classical-ISCO-energy}
\end{align}
then $\bar{\R}_{3}$ and $\bar{\R}_{4}$ coincide and yield the  classical ISCO radius 
\begin{align}
\bar{\mathcal{R}}_{\rm ISCO}= 3 \RS.
\label{classical-ISCO-radius}
\end{align}

\subsubsection{Analysis of the quartic equation in the quantum regime}\label{Sec:quartic-analysis-quantum}

We are now ready to address the resolution of Eq. \eqref{quartic-timelike} in the quantum scenario, where $k_1\neq 0$.

Let us first consider the case $E^2=1$. In this hypothesis, the quartic  \eqref{quartic-timelike} reduces to the quadratic equation
\begin{align}
\R^2 \RS  -2\RS ^2 \R +\frac{1}{2} k_1 \RS \LP^2   =0\,, \label{quadric-timelike}
\end{align}
which gives the exact solutions  
\begin{align}\label{R-plus-minus-energy-one}
\R_{\pm}=\RS \pm \frac{1}{2}\sqrt{4 \RS ^2-2k_1 \LP ^2}\,.
\end{align}
The corresponding values of the angular momentum follow from Eq. \eqref{L-2-and-E-timelike} and read as
\begin{align}
L_{\pm}^2=\RS \left(2\RS \pm \sqrt{4\RS ^2-2k_1 \LP ^2}  \right)\,,
\end{align}
while thanks to Eq. \eqref{second-derivative-potential-timelike} we obtain 
\begin{align}
V^{\prime \prime}_{\rm t}(\R_{\pm})&= -\frac{8 \RS \left[\RS \left(\sqrt{4 \RS^2-2 k_1  \LP ^2}\pm 2 \RS\right)\mp k_1 \LP ^2\right]}{\sqrt{\RS^2-\frac{k_1 \LP ^2}{2}} \left(\sqrt{4 \RS^2-2 k_1 \LP ^2}\pm 2 \RS\right)^4}.
\label{V-prime-prime-timelike-quantum-1}
\end{align}
It should be noted that the radii  $\R_{\pm}$ correspond to \emph{different}   envelopes of the effective potential as they are characterized by distinct  $L^2$.

If we exploit the approximation \eqref{Rs-bigger-than-lp}, which is valid within the EFT paradigm,  then from   Eq. \eqref{R-plus-minus-energy-one} we can write 
\begin{align}
\R_+ &= 2 \RS -(k_1/4) (\LP^2/\RS) + \OO\left(\LP^4/\RS^3\right),
\label{R-plus-quantum-approx}
\\
\R_- &= (k_1/4) (\LP^2/\RS) + \OO\left(\LP^4/\RS^3\right),    
\label{R-minus-quantum-approx}
\end{align}
the  solution \eqref{R-plus-quantum-approx} being in agreement with the classical result \eqref{bar-R-plus} in the limit $k_1=0$; moreover, upon  discarding higher-order corrections and recalling the condition \eqref{k1-O1},  Eq. \eqref{V-prime-prime-timelike-quantum-1} yields
\begin{align}
V^{\prime \prime}_{\rm t}(\R _+) & =-\frac{64\RS^6 \left(8\RS^2-3 k_1 \LP ^2\right)}{\left(k_1\LP ^2-8\RS^2\right)^4 (4\RS^2-k_1\LP ^2)}<0\,,
\label{V-prime-prime-4}
\\
V^{\prime \prime}_{\rm t}(\R _-)&= -\frac{64 \RS^6}{k_1^3 \LP ^6 (4\RS^2-k_1 \LP^2)}\,.
\label{V-prime-prime-3}
\end{align}

From the preceding equations, it is evident that in our hypotheses both $\R_+$ and $L^2_+$ remain positive regardless of the sign assumed by $k_1$, with the former being the maximum of the effective potential. This is not true for $\R_-$ and  $L_-^2$. In fact,   they  are both negative, and hence nonphysical, when $k_1$ is negative. For  $k_1>0$, $\R_{-}$ is positive and represents the maximum of the effective potential, since $V^{\prime \prime}_{\rm t}(\R_{-})<0$. Therefore, we can conclude that no SCO exists when $E^2 = 1$. This situation reflects the classical scenario, where the value $\bar{E}^2 =1$ gives the root  $\bar{\R}_+=2 \RS$ (cf. Eq. \eqref{bar-R-plus}), which corresponds to a radius  falling within the range of UCOs (see Eq. \eqref{classical-bound-1}).

At this stage, we can examine the general case with $E^2 \neq 1$. With this assumption,  Eq. \eqref{quartic-timelike} can be expressed in the standard quartic form:
\begin{subequations}\label{eq:quartic_final}
   \begin{align}\label{eq:quartic_final1}
\R^4+a_3 \R^3+a_2 \R^2+a_1 \R +a_0=0\,,
\end{align}
\text{with}
\begin{align}\label{eq:quartic_final2}
    a_3&=\frac{\left({3 E^2}/{2}-2\right) \RS}{1-E^2}\,,\notag \\
    a_2&=\frac{ \RS^2 }{1-E^2}\,,\notag \\
    a_1&=\frac{\left(3 E^2-{7}/{2}\right) k_1 \LP ^2 \RS}{2 (1-E^2)} \,,\notag \\
   a_0&= \frac{3 k_1 \LP ^2 \RS^2}{2}\,.
\end{align} 
\end{subequations}

In the literature, there exist various methodologies to solve  a quartic exactly  \cite{abramowitz1968handbook}. However, these approaches usually envisage  a long procedure and  the ensuing roots  involve lengthy  expressions. For this reason, we now propose a simple analytical technique for  finding the approximated form of the solutions of the quartic \eqref{eq:quartic_final}, which  thus assume
a readable and ready-to-use form. Following the principles of the EFT scheme, we suppose that  the roots  of Eq. \eqref{eq:quartic_final} can be expanded as  power series in   $(\LP/\RS)^n$, where the leading-order term  is represented by the classical solution and $n$ is a positive real number to be determined shortly. Accordingly, we have
\begin{subequations}
\begin{align}
  \mathcal{R}_{1} &=\bar{\mathcal{R}}_1+f_1 \left(\LP/\RS \right)^n + f_2 \left(\LP / \RS\right)^{2n} + \OO \left(\LP /\RS \right)^{3n}\,,\\
  \mathcal{R}_2 &=\bar{\mathcal{R}}_2+ f_3 \left(\LP/\RS \right)^n + f_4 \left(\LP / \RS\right)^{2n} + \OO \left(\LP /\RS \right)^{3n}\,,\\
   \mathcal{R}_3 &=\bar{\mathcal{R}}_3+ f_5 \left(\LP/\RS \right)^n + f_6 \left(\LP / \RS\right)^{2n}+ \OO \left(\LP /\RS \right)^{3n}\,,\\
    \mathcal{R}_4 &=\bar{\mathcal{R}}_4+ f_7 \left(\LP/\RS \right)^n + f_8 \left(\LP / \RS\right)^{2n}  + \OO \left(\LP /\RS \right)^{3n}\,,
\end{align}   
\end{subequations}
with $\bar{\mathcal{R}}_1=\bar{\mathcal{R}}_2=0$ (i.e., $\mathcal{R}_{1}$ and $\mathcal{R}_{2}$
are fully quantum radii), while $\bar{\mathcal{R}}_3$ and $\bar{\mathcal{R}}_4$ are given in Eq. \eqref{bar-R-3-and-4}.  Bearing in mind the coefficients~\eqref{eq:quartic_final2}, a self-consistent perturbative solution of the quartic  can be achieved when $n=1$. Then, the factors $f_i$ can be determined by substituting $\mathcal{R}_i$ back into Eq.~\eqref{eq:quartic_final}. For example, by plugging $\mathcal{R}_1$  into Eq.~\eqref{eq:quartic_final}, we have 
\begin{align}
&\left(\frac{f_1^2 }{1-E^2}+\frac{3 k_1 \RS^2}{2} \right) \LP ^2+ \left[f_1 \frac{ \left(8-6 E^2\right) f_1^2/\RS ^2+\left(7-6 E^2\right) k_1-8 f_2 /\RS}{4 \left(E^2-1\right)} \right]\LP^3
\nonumber \\
&+\OO \left(\LP^4 \right)=0\,.
\end{align}
This equation requires that the terms proportional to $\LP^2$ and $\LP^3$  be zero separately, i.e., 
\begin{align}
    0&=\frac{f_1^2}{1-E^2}+\frac{3 k_1 \RS^2}{2}\,,\\
    0&=\frac{f_1  \left[\left(8-6 E^2\right) f_1^2/\RS^2+\left(7-6 E^2\right) k_1-8 f_2 /\RS\right]}{4 \left(E^2-1\right)}\,,
\end{align}
 which yield
\begin{align}
    f_1&= \pm   \sqrt{\frac{3 k_1}{2}(E^2 -1)} \, \RS \,,\\
    f_2&=\frac{-9 E^4 +15 E^2 -5 }{8 }k_1\RS\,.
\end{align}
Similarly,  considering $\R_2$ we obtain  
\begin{align}
    f_3&=\pm \sqrt{\frac{3 k_1}{2}(E^2 -1)} \, \RS  \,,\\
    f_4&=f_2\,.
\end{align}
To be specific, in order to let $\mathcal{R}_1$ and $\mathcal{R}_2$ represent distinct solutions, we choose different values for $f_1$ and $f_3$, i.e.,
\begin{align}
f_1=-f_3=\sqrt{\frac{3 k_1}{2}(E^2 -1)} \, \RS \,,
\end{align}
which means that $\mathcal{R}_{1}$  and   $\mathcal{R}_{2}$ take the form
\begin{subequations}
\label{eq:sols-quartic}
\begin{align}\label{eq:sols-quartic1}
    \mathcal{R}_{1} &=\sqrt{\frac{3 k_1}{2}(E^2 -1)} \, \LP -\left(\frac{9 E^4 -15 E^2 +5 }{8 }\right) k_1 \LP^2/\RS  + \OO (\LP^3 /\RS ^2)\,,\\
  \mathcal{R}_2 &=-\sqrt{\frac{3 k_1}{2}(E^2 -1)} \, \LP -\left(\frac{9 E^4 -15 E^2 +5 }{8 } \right)k_1 \LP^2/\RS  + \OO (\LP^3 /\RS ^2)\,.
\end{align} 
\text{By employing the same method, we obtain the other two roots $\mathcal{R}_{3}$ and $\mathcal{R}_{4}$ for the quartic equation.}\\
\text{For $E^2 \neq 8/9$, these  are given by}
\begin{align}
\mathcal{R}_{3}^{\rm  case1} &=\bar{\mathcal{R}}_3+\left[\frac{9 E^4 -15 E^2 +5 }{8 }-\frac{\left(27 E^6-57 E^4+27 E^2+4\right)  }{8 E \sqrt{9 E^2-8}  }\right] k_1 \LP ^2/\RS 
\nonumber \\   
&+ \OO (\LP^3 /\RS ^2)\,, \label{eq:sols-quartic3}
\\ 
\mathcal{R}_4 ^{\rm  case1} &=\bar{\mathcal{R}}_4+\left[\frac{9 E^4 -15 E^2 +5 }{8 }+\frac{\left(27 E^6-57 E^4+27 E^2+4\right)  }{8 E \sqrt{9 E^2-8}  }\right] k_1 \LP ^2/\RS  
\nonumber \\ 
&+ \OO (\LP^3 /\RS ^2) \,,
\label{eq:sols-quartic4}
\end{align}
\text{while for $E^2 = 8/9$, we have}
\begin{align}\label{eq:sols-quartic5}
    \mathcal{R}_{3} ^{\rm  case2} &=3\RS-\frac{1}{2} \sqrt{\frac{13k_1}{3}} \, \LP -\left(\frac{11 k_1}{72}\right)  \LP ^2/\RS + \OO (\LP^3 /\RS ^2)\,,\\ \label{eq:sols-quartic6}
  \mathcal{R}_4 ^{\rm  case2} &=3\RS+\frac{1}{2} \sqrt{\frac{13k_1}{3}} \, \LP -\left(\frac{11 k_1 }{72} \right) \LP ^2/\RS+ \OO (\LP^3 /\RS ^2) \,.
\end{align}
\end{subequations}

It is possible to unify these two sets of solutions into a single one, yielding
\begin{align}\label{eq:sols-quartic-combine1}
    \mathcal{R}_{3} &=\frac{\RS}{4(E^2-1)}\sqrt{E^2(9E^2-8)+(4-27E^8+84E^6-84E^4+23E^2){k_1 \LP^2}/{\RS^2}} \notag \\
    &\qquad +\frac{(3E^2-4)\RS}{4(E^2-1)}+\frac{(9 E^4 -15 E^2 +5 )k_1 \LP ^2}{8 \RS}+\OO (\LP^3 /\RS ^2)\,,\\ \label{eq:sols-quartic-combine2}
  \mathcal{R}_4 &=-\frac{\RS}{4(E^2-1)}\sqrt{E^2(9E^2-8)+(4-27E^8+84E^6-84E^4+23E^2){k_1 \LP^2}/{\RS^2}} \notag \\
    &\qquad +\frac{(3E^2-4)\RS}{4(E^2-1)}+\frac{(9 E^4 -15 E^2 +5 )k_1 \LP ^2}{8 \RS}+\OO (\LP^3 /\RS ^2)\,.
\end{align}

\begin{table}
    \begin{tabular}{|c|c|c|c|c|}
     \multicolumn{5}{c}{$k_1=1\,,E=10\,, \LP/\RS=10^{-4}$} \\
     \hhline{|=====|}
    Solution  &$\R_1 /\RS$ &$\R_2 /\RS $& $\R_3 /\RS$  &$\R_4 /\RS$  \\
    \hline
    Analytic solution&0.00111 &-0.00133&1.50168 & -0.00651 \\
    \hline
    Numerical solution&0.00111 &-0.00137&  1.50168&-0.00649  \\
    \hline
    $V_{\rm t}^{\prime \prime}(\mathcal{R}_i) \RS^2$&$6.05177\times 10^8$ &$-3.59837\times 10^8$ &-132.001 & $-3.59127\times 10^6$\\
    \hline
    Sign of $L^2$& -&-& +&-  \\
    \hline
    \end{tabular}
    \\ \vspace{2mm}
    \begin{tabular}{|c|c|c|c|c|}
     \multicolumn{5}{c}{$k_1=1\,,E=2\,, \LP/\RS=10^{-4}$} \\
     \hhline{|=====|}
    Solution  &$\R_1 /\RS$ &$\R_2 /\RS $& $\R_3 /\RS$  &$\R_4 /\RS$  \\
    \hline
    Analytic solution&0.00021 &-0.00021&1.54858 & -0.21525 \\
    \hline
    Numerical solution&0.00021 &-0.00021& 1.54858&-0.21525  \\
    \hline
     $V_{\rm t}^{\prime \prime}(\mathcal{R}_i) \RS^2$&$1.04941\times 10^{11}$ &$-1.04572\times 10^{11}$ &-4.02223 & $-93.9781$\\
    \hline
    Sign of $L^2$& -&-& +&-  \\
    \hline
    \end{tabular}
    \\ \vspace{2mm}
    \begin{tabular}{|c|c|c|c|c|}
     \multicolumn{5}{c}{$k_1=1\,,E=1.01\,, \LP/\RS=10^{-4}$} \\
     \hhline{|=====|}
    Solution  &$\R_1 /\RS$ &$\R_2 /\RS $& $\R_3 /\RS$  &$\R_4 /\RS$  \\
    \hline
    Analytic solution&0.00002 &-0.00002&1.96342 & -25.33900 \\
    \hline
    Numerical solution&0.00002 &-0.00002&1.96342 & -25.33900  \\
    \hline
    $V_{\rm t}^{\prime \prime}(\mathcal{R}_i) \RS^2$&$1.90925\times 10^{14}$ &$-1.91106\times 10^{14}$ &-0.741159 & $-0.00032$\\
    \hline
    Sign of $L^2$& +&+& +&-  \\
    \hline
    \end{tabular}
    \\ \vspace{2mm}
    \begin{tabular}{|c|c|c|c|c|}
     \multicolumn{5}{c}{$k_1=1\,,E=0.99\,, \LP/\RS=10^{-4}$} \\
     \hhline{|=====|}
    Solution  &$\R_1 /\RS$ &$\R_2 /\RS $& $\R_3 /\RS$  &$\R_4 /\RS$  \\
    \hline
    Analytic solution&0.00002i &-0.00002i& 2.04428 & 24.58130 \\
    \hline
    Numerical solution&0.00002i &-0.00002i&2.04428 & 24.58130  \\
    \hline
     $V_{\rm t}^{\prime \prime}(\mathcal{R}_i) \RS^2$&Complex &Complex &-0.10277 & $0.00003$\\
    \hline
    Sign of $L^2$& Complex& Complex& +&+  \\
    \hline
    \end{tabular}
   \\ \vspace{2mm}
    \begin{tabular}{|c|c|c|c|c|}
     \multicolumn{5}{c}{$k_1=1\,,E=\sqrt{8/9}\,, \LP/\RS=10^{-4}$} \\
     \hhline{|=====|}
    Solution  &$\R_1 /\RS$ &$\R_2 /\RS $& $\R_3 /\RS$  &$\R_4 /\RS$  \\
    \hline
    Analytic solution&0.00004i &-0.00004i& 2.99990 & 3.00010 \\
    \hline
    Numerical solution&0.00004i &-0.00004i& 2.99990 & 3.00010  \\
    \hline
     $V_{\rm t}^{\prime \prime}(\mathcal{R}_i) \RS^2$&Complex &Complex &$-1.28515\times 10^{-6}$ & $1.28481\times 10^{-6}$\\
    \hline
    Sign of $L^2$& Complex& Complex& +&+  \\
    \hline
    \end{tabular}
    \\ \vspace{2mm}
    \begin{tabular}{|c|c|c|c|c|}
     \multicolumn{5}{c}{$k_1=1\,,E=0.5\,, \LP/\RS=10^{-4}$} \\
     \hhline{|=====|}
    Solution  &$\R_1 /\RS$ &$\R_2 /\RS $& $\R_3 /\RS$  &$\R_4 /\RS$  \\
    \hline
    Analytic solution&0.00011i &-0.00011i& 1.08333-0.39965i & 1.08333-0.39965i \\
    \hline
    Numerical solution &0.00011i &-0.00011i& 1.08333-0.39965i & 1.08333-0.39965i  \\
    \hline
     $V_{\rm t}^{\prime \prime}(\mathcal{R}_i) \RS^2$&Complex &Complex &Complex & Complex\\
    \hline
    Sign of $L^2$& Complex& Complex& Complex&Complex  \\
    \hline
    \end{tabular}
    \caption{Roots (in units of $\RS$) of the quartic \eqref{eq:quartic_final} obtained analytically via our method and numerically for  $k_1=1$ and  various choices of the energy $E$;  the values of $V_{\rm t}^{\prime \prime}(\mathcal{R}_i)$ and $L^2$, which follow from Eqs. \eqref{second-derivative-potential-timelike}  and  \eqref{L-2-and-E-timelike}, respectively, are also provided. The agreement between our approach and the numerics  is evident.}
    \label{table1}
\end{table}

\begin{table}
\centering
\resizebox{14cm}{!}{%
    \begin{tabular}{|c|c|c|c|c|}
     \multicolumn{5}{c}{$k_1=-1\,,E=10\,, \LP/\RS=10^{-4}$} \\
     \hhline{|=====|}
    Solution  &$\R_1 /\RS$ &$\R_2 /\RS $& $\R_3 /\RS$  &$\R_4 /\RS$  \\
    \hline
    Analytic solution&0.00011+0.00122i &0.00011-0.00122i&1.50168 & -0.00695 \\
    \hline
    Numerical solution&0.00010+0.00120i &0.00010-0.00120i&1.50168 & -0.00693  \\
    \hline
     $V_{\rm t}^{\prime \prime}(\mathcal{R}_i) \RS^2$&Complex &Complex &-132.001 & $-2.95082\times 10^6$\\
    \hline
    Sign of $L^2$& Complex& Complex& +&-\\
    \hline
    \end{tabular}}
    \\  \vspace{2mm}
\resizebox{0.6\textwidth}{!}{%
  \begin{tabular}{|c|c|c|c|c|}
     \multicolumn{5}{c}{$k_1=-1\,,E=2\,, \LP/\RS=10^{-4}$} \\
     \hhline{|=====|}
    Solution  &$\R_1 /\RS$ &$\R_2 /\RS $& $\R_3 /\RS$  &$\R_4 /\RS$  \\
    \hline
    Analytic solution&0.00021i &-0.00021i&1.54858 &-0.21525 \\
    \hline
    Numerical solution&0.00021i &-0.00021i& 1.54858&-0.21525 \\
    \hline
    $V_{\rm t}^{\prime \prime}(\mathcal{R}_i) \RS^2$&Complex &Complex &-4.02223 & -93.9775\\
    \hline
   Sign of $L^2$ & Complex & Complex & + & - \\
     \hline
    \end{tabular}}
    \\ \vspace{2mm}
   \resizebox{0.6\textwidth}{!}{%
   \begin{tabular}{|c|c|c|c|c|}
     \multicolumn{5}{c}{$k_1=-1\,,E=1.01\,, \LP/\RS=10^{-4}$} \\
     \hhline{|=====|}
    Solution  &$\R_1 /\RS$ &$\R_2 /\RS $& $\R_3 /\RS$  &$\R_4 /\RS$  \\
    \hline
    Analytic solution&0.00002i &-0.00002i&1.96342 & -25.33900 \\
    \hline
    Numerical solution&0.00002i &-0.00002i&1.96342 & -25.33900  \\
    \hline
     $V_{\rm t}^{\prime \prime}(\mathcal{R}_i) \RS^2$&Complex &Complex &-0.147759 & -0.00003\\
    \hline
      Sign of $L^2$& Complex& Complex& +&-\\
    \hline
    \end{tabular}}
    \\ \vspace{2mm}
   \resizebox{11.5cm}{!}{%
   \begin{tabular}{|c|c|c|c|c|}
     \multicolumn{5}{c}{$k_1=-1\,,E=0.99\,, \LP/\RS=10^{-4}$} \\
     \hhline{|=====|}
    Solution  &$\R_1 /\RS$ &$\R_2 /\RS $& $\R_3 /\RS$  &$\R_4 /\RS$  \\
    \hline
    Analytic solution&0.00002 &-0.00002& 2.04428 & 24.58130 \\
    \hline
    Numerical solution&0.00002 &-0.00002&2.04428 & 24.58130  \\
    \hline
     $V_{\rm t}^{\prime \prime}(\mathcal{R}_i) \RS^2$&$1.94006\times 10^{14}$ &$-1.93799\times 10^{14}$ &-0.102766 &0.00003\\
    \hline
     Sign of $L^2$& -& - & + &+\\
    \hline
    \end{tabular}}
    \\ \vspace{2mm}
      \resizebox{0.69\textwidth}{!}{%
      \begin{tabular}{|c|c|c|c|c|}
     \multicolumn{5}{c}{$k_1=-1\,,E=\sqrt{8/9}+0.00001\,, \LP/\RS=10^{-4}$} \\
     \hhline{|=====|}
    Solution  &$\R_1 /\RS$ &$\R_2 /\RS $& $\R_3 /\RS$  &$\R_4 /\RS$  \\
    \hline
    Analytic solution&0.00004 &-0.00004& 2.97274 & 3.02802 \\
    \hline
    Numerical solution&0.00004 &-0.00004& 2.97274 & 3.02802  \\
    \hline
     $V_{\rm t}^{\prime \prime}(\mathcal{R}_i) \RS^2$&$1.47047\times 10^{13}$ &$-1.46966\times 10^{13}$ &-0.00035 &0.00033\\
    \hline
    Sign of $L^2$&-& -& +&+\\
    \hline
    \end{tabular}}
    \\ \vspace{2mm}
    \resizebox{12.6cm}{!}{%
    \begin{tabular}{|c|c|c|c|c|}
     \multicolumn{5}{c}{$k_1=-1\,,E=\sqrt{8/9}\,, \LP/\RS=10^{-4}$} \\
     \hhline{|=====|}
    Solution  &$\R_1 /\RS$ &$\R_2 /\RS $& $\R_3 /\RS$  &$\R_4 /\RS$  \\
    \hline
    Analytic solution&0.00004 &-0.00004& 3-0.00010i & 3+0.00010i \\
    \hline
    Numerical solution&0.00004 &-0.00004& 3-0.00010i & 3+0.00010i  \\
    \hline
    $V_{\rm t}^{\prime \prime}(\mathcal{R}_i) \RS^2$&$1.47010\times 10^{13}$ &$-1.46929\times 10^{13}$ &Complex &Complex\\
    \hline
     Sign of $L^2$& -& - & Complex &Complex\\
    \hline
    \end{tabular}}
    \\ \vspace{2mm}
  \resizebox{14.5cm}{!}{%
  \begin{tabular}{|c|c|c|c|c|}
     \multicolumn{5}{c}{$k_1=-1\,,E=0.5\,, \LP/\RS=10^{-4}$} \\
    \hhline{|=====|}
    Solution  &$\R_1 /\RS$ &$\R_2 /\RS $& $\R_3 /\RS$  &$\R_4 /\RS$  \\
   \hline
   Analytic solution&0.00011 &-0.00011& 1.08333-0.39965i & 1.08333-0.39965i \\
    \hline
    Numerical solution &0.00011 &-0.00011& 1.08333-0.39965i & 1.08333-0.39965i  \\
   \hline
    $V_{\rm t}^{\prime \prime}(\mathcal{R}_i) \RS^2$&$8.38127\times 10^{11}$ &$-8.37978\times 10^{11}$ &Complex &Complex\\
    \hline
   Sign of $L^2$&-& -& Complex&Complex  \\
  \hline
    \end{tabular}}
    \caption{Roots (in units of $\RS$) of the quartic \eqref{eq:quartic_final} for  $k_1=-1$ and different values of $E$. Our analytical method and the numerical approach agree with high accuracy.}
    \label{table2}
\end{table}

It is easy to check that the analytic solutions \eqref{eq:sols-quartic}--\eqref{eq:sols-quartic-combine2} provided by our approach are such that the Vi\`ete formulas for  quartic equations \cite{abramowitz1968handbook} are satisfied modulo higher-order terms  in $\LP$, as we obtain
\begin{subequations}\label{eq:Viete formulas}
    \begin{align}\label{eq:Viete formulas1}
    &\sum _{i=1} ^{4} \R _i = -a_3+\OO\left(\LP^3/\RS^2\right) \,, \\\label{eq:Viete formulas2}
    & \sum _{i,j=1} ^{4} \R_i \R_j= a_2 +\OO\left(\LP^4/\RS^2\right)\,, \qquad  \quad \quad \quad \text{for } i\neq j\,,\\
     & \sum _{i,j,k=1} ^{4} \R_i \R_j \R_k= -a_1 +\OO\left(\LP^4/\RS\right)\,,\qquad \;  \, \text{for } i\neq j \neq k \, ,\\
    &\R_1 \R_2 \R_3 \R_4=a_0 +\OO\left(\LP^4\right)\,.
\end{align}
\end{subequations}
Moreover, we find that our  formulas  align closely with the numerical solutions, as evidenced in Tables \ref{table1} and \ref{table2}. As can  be inferred from  the values of $V_{\rm t}^{\prime \prime}(\mathcal{R}_i)$ ($i=1,2,3,4$) reported in the tables,  we have also verified that $\R_{3}$ is the maximum while $\R_{4}$ the minimum of the effective potential when $E^2_{\rm ISCO}<E^2<1$, where the quantum ISCO energy $E^2_{\rm ISCO}$ will be derived in Eq. \eqref{E-squared-ISCO-quantum} below. 

It is worth noticing that, although the solutions \eqref{eq:sols-quartic}--\eqref{eq:sols-quartic-combine2} have been derived by supposing $E \neq 1$, some of them admit a continuous extension when $E=1$. Indeed, up to higher-order corrections, we have that $\R_1^{(E=1)}+\R_2^{(E=1)}= \R_-$, and $\R_3^{(E=1)} =\R_+ $ like in the classical case (see Eq. \eqref{bar-R-plus}); on the other hand, $\R_4$ cannot be extended at $E=1$, since its classical piece $\bar{\R}_4$ is not defined in this limit.

At this stage, some remarks should be made. Following   EFT principles, quantum contributions to a physical quantity should not become more important than the classical ones. Since quantum terms  in $\R_{i}$ ($i=1,2,3,4$) depend on $E$, some constraints on this constant should be taken into account.  While mathematically there may be no upper bound on $E$, physically  the energy  cannot be arbitrarily large, otherwise one can no longer neglect  the particle back-reaction on the geometry. This is true also in our model, as $\R_{3,4}$ appear in  the form of a classical piece $\bar{\R}_{3,4}$ plus a tiny quantum factor, if
\begin{align} \label{energy-bound}
    E \ll \RS/\LP,
\end{align}
or, equivalently,
\begin{align}
    E \ll M /\MP, \label{energy-bound-3}
\end{align}
which prevent the energy from attaining huge values in view of  Eqs. \eqref{Rs-bigger-than-lp} and \eqref{M-bigger-M-Planck}. Therefore, the above equations suggest that $\RS/\LP$, or equivalently $M /\MP$,   can be interpreted as a cutoff energy scale of our model.

In the classical scenario with $k_1=0$, the attainment of the ISCO occurs precisely when $\bar{E}_{\rm ISCO}^2=8/9$ (cf. Eq. \eqref{classical-ISCO-energy}). The classical ISCO radius can be determined by equating Eq.~\eqref{eq:sols-quartic5} to Eq.~\eqref{eq:sols-quartic6} in the limit $k_1=0$,  which indeed yields the value \eqref{classical-ISCO-radius}.  
This criterion is no longer sustained in the   quantum realm, as Eq.~\eqref{eq:sols-quartic5} differs from Eq.~\eqref{eq:sols-quartic6} for $k_1 \neq 0$.  For this reason, we first find the ISCO energy   by equating Eq.~\eqref{eq:sols-quartic-combine1} to Eq.~\eqref{eq:sols-quartic-combine2},  which gives
\begin{align}\label{eq:isco-1}
    E^2 \left(9 E^2-8\right) \RS^2+\left(-27 E^8+84 E^6-84 E^4+23 E^2+4\right) k_1 \LP^2=0\,.
\end{align}
Not all of the roots of Eq.~\eqref{eq:isco-1} are physically relevant, since we are interested in the solution which slightly deviates from the corresponding classical one. Therefore, we exploit again our method and assume that $ E^2_{\rm ISCO}$ admits the expansion
\begin{align}
    E^2_{\rm ISCO}=\bar{E}_{\rm ISCO}^2+f_9 \left(\LP/\RS\right)+f_{10} \left(\LP^2/\RS ^2 \right)+\OO \left(\LP^3 /\RS ^3 \right)\,,
\end{align}
where the classical value $\bar{E}_{\rm ISCO}^2$ has been given in Eq. \eqref{classical-ISCO-energy}. Upon substituting the above expression  back into Eq.~\eqref{eq:isco-1} and imposing the condition that terms proportional to $\LP$ and $\LP^2$ vanish independently,  we obtain 
\begin{align}
    f_9&=0\,,\\
    f_{10}&=-\frac{13 k_1}{486}\,,
\end{align}
and hence we can write 
\begin{align}
E^2_{\rm ISCO}=8/9 -\left(\frac{13 k_1}{486}\right)\LP^2/\RS ^2+\OO (\LP^3 /\RS ^3)\,,
\label{E-squared-ISCO-quantum}
\end{align}
which,  jointly with either Eq.~\eqref{eq:sols-quartic-combine1} or Eq. \eqref{eq:sols-quartic-combine2}, permits obtaining the quantum corrected ISCO radius 
\begin{align}
\R _{\rm ISCO}=3 \RS- \left(\frac{25 k_1 }{36 }\right) \LP^2/ \RS +\OO \left(\LP^3 /\RS ^2\right)\,.
\label{R-ISCO-new-method}
\end{align}
From the above relations, we find that the binding energy $E_B$ per unit rest mass of the ISCO is 
\begin{align}
 E_B=1- E_{\rm ISCO} =1-\sqrt{8/9} \left[1-\left(\frac{13 k_1}{864}\right)\LP^2/\RS ^2+\OO (\LP^3 /\RS ^3)\right].
\end{align}

Similarly to the classical scenario, $ V_{\rm t}^{\prime \prime}(\mathcal{R}_{\rm ISCO})$ vanishes up to higher-order corrections. In fact, by means of  Eqs.~\eqref{E-squared-ISCO-quantum} and \eqref{R-ISCO-new-method}, we find from  Eq.~\eqref{second-derivative-potential-timelike}
\begin{align}
    \RS^2 \, V_{\rm t}^{\prime \prime}(\mathcal{R}_{\rm ISCO})= -\frac{k_1 \LP^2}{972\,\RS^2}+\OO(\LP ^4/ \RS^4)\,, 
\end{align}
which, when using the same numerical values as in Tables \ref{table1} and \ref{table2} for $\LP/\RS $ and $ k_1 $, gives  $\RS^2 \, V_{\rm t}^{\prime \prime}(\mathcal{R}_{\rm ISCO}) \sim 10^{-11}$.

In classical GR, the radii of  UCOs and SCOs  are subject to the conditions  \eqref{classical-bound-1} and \eqref{classical-bound-2}, respectively. Analogous  constraints exist also in the quantum regime. In fact,  as we have shown above,  when $E^2_{\rm ISCO}\leq E^2 <1$, SCOs have radius $\R_4 \geq \R _{\rm ISCO}$. On the other hand, bearing in mind the classical result \eqref{limit-classic-unstable}, the  radius $\R _{\rm IUCO}$ of the innermost UCO (located outside the black hole) could be found by solving Eq.~\eqref{eq:quartic_final} in the limit $E^2\gg 1$ (while still complying with the requirement  \eqref{energy-bound})\footnote{Taking the limit $E^2 \gg 1$ in the solution \eqref{eq:sols-quartic-combine1} will not lead to the correct result, as this equation is not reliable when $E^2 $ attains unboundedly large values.}. Physically, $\R _{\rm IUCO}$ should be equal to the radius of photon sphere, i.e., the unstable circular  trajectory followed by  photons. This actually occurs in our model, as  when $E\gg 1$ Eq.~\eqref{eq:quartic_final} reduces to  
\begin{align}\label{eq:quartic_largeE}
\R^4-\frac{3 \RS}{2} \R^3-\frac{3}{2}k_1 \RS \LP ^2  \R +\frac{3}{2}k_1 \RS ^2 \LP ^2 =0\,,
\end{align}
which takes \emph{exactly} the same form as the equation pertaining to the photon sphere radius to  be derived in  Eq. \eqref{eq:r-sphere-1} below. Therefore, a situation mirroring the classical scenario is valid also for UCOs, since  
\begin{align}
\R _{\rm IUCO}<\R_3 < \R_{\rm ISCO}, \label{R-IUCO-R-3-R-ISCO}
\end{align}
with $E^2 > E^2_{\rm ISCO}$ and $\R _{\rm IUCO} \equiv \mathcal{R}_{\rm ps 4} $, see Eq. \eqref{R-ps-quantum-4} below.

Although the EFT scheme should be used with caution at scales of the order  of the Planck length,  it can still offer a first rough account of quantum phenomena, giving rise to interesting scenarios. Indeed, one intriguing feature that distinguishes our model from classical general relativity is its prediction of (at least\footnote{The scenario with positive $k_1$ admits at least two disconnected SCO regions due to the  presence of the two null hypersurfaces with radii $\tilde{r}_1$ and $\tilde{r}_2$ (see Eqs. \eqref{solution-r-tilde-1} and \eqref{solution-r-tilde-2}).}) two disconnected SCO regions when $k_1>0$. In principle, bearing in mind Eq. \eqref{eq:sols-quartic1},  there should exist SCOs with radius $\R_1$ residing in the black hole whenever  $E^2>1>E^2_{\rm ISCO}$. However, our analysis shows that this happens  if $E \gtrsim 1$, a situation which yields $\R_1 \lesssim \LP$. This can be inferred from Table \ref{table1}, where it is evident that  when $E=1.01$ we have $\R_1 \sim 10^{-1} \, \LP$, with  $V_{\rm t}^{\prime \prime}(\mathcal{R}_1)  >0$  and $L^2$ attaining a positive value; on the other hand, the solutions $\R_1$ with $E=2$ and $E=10$ lead to a negative  $L^2$, and hence must be regarded as nonphysical. Since $\R_1<\tilde{r}_1$ (see Eq. \eqref{solution-r-tilde-1}), the corresponding SCO trajectories cannot be observed from outside the black hole, and hence  the potential physical inconsistencies deriving from the fact that $\R_1 $ can become smaller than $\LP$ when $E \gtrsim 1$ present no issues. 

Due to the presence of these inner SCO orbits, we can conclude that the pattern with positive $k_1$ features the presence of an ISCO located \emph{deeply} inside the black hole  (as first proved in Ref. \cite{Battista2023}). Therefore,  we can regard Eq. \eqref{R-ISCO-new-method} as the quantum corrected version of the classical ISCO radius \eqref{classical-ISCO-radius}. This should be regarded as the  \qm{physical} ISCO, and hence hereafter we will refer to  Eq. \eqref{R-ISCO-new-method} as simply the ISCO. 

An equivalent way to prove that the regime with $k_1>0$ allows for (at least) two disconnected SCO domains is as follows. If the SCO regions were connected, a SCO having radius $\tilde{r}_1$ would be viable. Therefore,  the condition $V^\prime_{\rm t} (\tilde{r}_1)=0$  would bring about the constraint 
\begin{align}\label{eq:SCO_contraint}
6 k_1 \left[\left(2 E^2-5\right) \RS^2+2 L^2\right]+4 \RS^2 \left(L^2+\RS^2\right)=0\,.
 \end{align}
Due to the relations \eqref{k1-O1} and \eqref{Rs-bigger-than-lp}, the above equation  cannot be satisfied, and hence there exists no SCO with radius $\tilde{r}_1$.  This is of crucial physical importance: the null hypersurface at $\tilde{r}_1$ cannot serve as the site of a SCO, as such a scenario would imply that massive particles could seemingly escape from the black hole. 

When $k_1 <0 $,  both SCOs and UCOs are located outside the black hole, as they lie beyond the null hypersurface having radius $\tilde{r}_3$ (see Fig. \ref{Fig-k1-negative}).  Indeed, in this case  the solution $\R_1$, which, when $E^2<1$,  can be either of the order of $\LP$ or even smaller, yield a negative value of  $L^2$ and hence is nonphysical (see   table \ref{table2}).

\subsection{Massive particles trajectories }\label{Sec:trajectory-timelike}

Since the effective potential \eqref{effective-potential-timelike}  has, apart from  $L$,  an explicit dependence on the energy $ \mathscr{E}$ and the factor $k_1$, the ensuing dynamics  can manifest intriguing features in the quantum domain.  In the classical scenario where $k_1=0$, the effective potential $V_{\rm t}^{k_1=0}(r)$ is uniquely determined once the angular momentum $L$ is fixed. Consequently, if there exists an UCO, a particle will fall into the black hole if its energy exceeds the value of the potential  evaluated at the UCO radius,  hereafter denoted with $\R_{\rm UCO}$. This relation remains valid in the quantum realm as well, albeit $\R_{\rm UCO}$  receives an additional contribution  from  $k_1$ (notice that $\R_{\rm UCO}$ coincides with the physically meaningful solutions  $\R_3$, see Eq. \eqref{eq:sols-quartic-combine1}). Even though this quantum  correction minimally affects the value of $\R_{\rm UCO}$, it can lead to entirely different particle orbits even if identical initial conditions are chosen. This interesting phenomenon will be  shown explicitly in this section. 

The  spatial orbits of massive particles  can be easily derived starting from Eqs. \eqref{geod-timelike-one-loop} and \eqref{effective-potential-timelike}, and then dividing the obtained formulas by $\dot{\phi}= L/r^2$ (which stems from the rotational invariance of the geometry  \eqref{Schwarzschild_metric_quantum-standard}). In this way,  we find  that the equation governing the  motion  of massive bodies on the equatorial plane $\theta =\pi/2$ is  
\begin{align}\label{eq:massive_particle_orbit}
\left( \frac{\dd r }{\dd \phi }\right)^2 +\left(1-\frac{\RS}{r}\right)\left( r^2+\frac{r^4}{L^2}\right)+\frac{k_1  \RS \LP^2 \,r}{L^2}\left(\frac{3}{2}-\mathscr{E}^2\right)=\frac{r^4 \mathscr{E}^2}{L^2}.
\end{align}
\begin{figure}
    \centering
    \begin{subfigure}[t]{0.45\textwidth}
   \includegraphics[width=7.0cm]{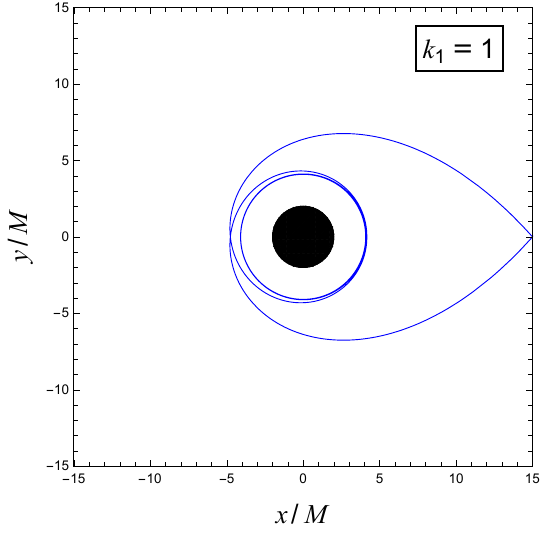}
  \end{subfigure} 
    \begin{subfigure}[t]{0.45\textwidth}
      \includegraphics[width=7.0cm]{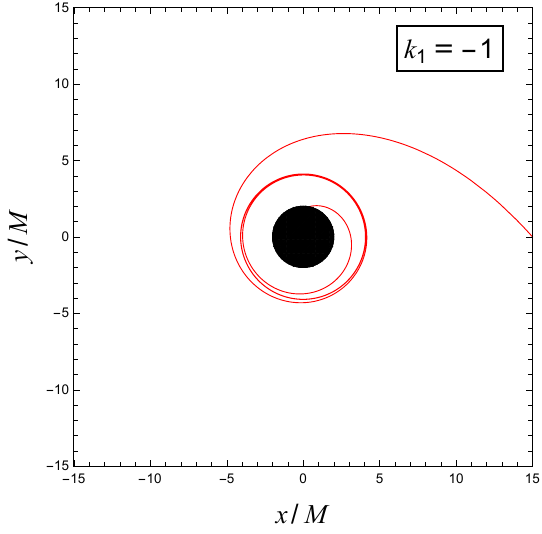}
  \end{subfigure}  
    \caption{Massive particle trajectories beginning at $x_0=15M, y_0=0$, with $ \mathscr{E}=0.99$ and $L=3.918712113126 M$ (we set $G=1$).  Although the initial conditions are the same, two distinct orbits are obtained: the body spirals  into the black hole (represented by a black disk) or returns almost to its starting point, depending on whether  $k_1=1$ or $k_1=-1$, respectively. For both situations, we have $\LP/\RS=10^{-4}$, which yields $\R _{\rm UCO} ^{k_1=1} = 2.044284344\RS$, $\R _{\rm UCO} ^{k_1=-1} = 2.044284345\RS$, and $V_{\rm t} (\R _{\rm UCO} ^{k_1=1})> \mathscr{E} > V_{\rm t} (\R _{\rm UCO} ^{k_1=-1})$. }
    \label{fig:trajectorytimelike2}
\end{figure}
\begin{figure}
    \centering
    \begin{subfigure}[t]{0.45\textwidth}
   \includegraphics[width=7.0cm]{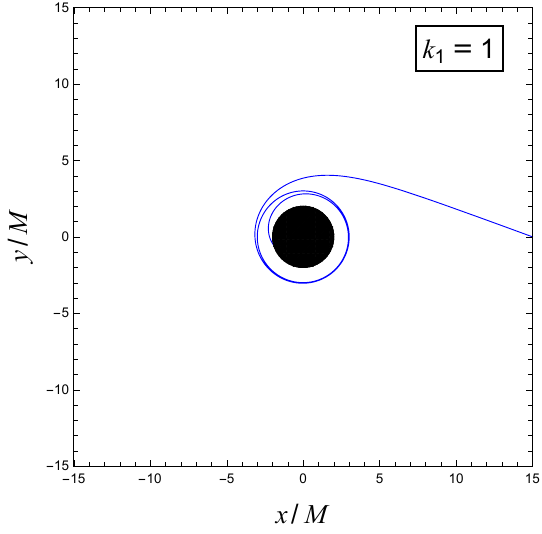}
  \end{subfigure} 
    \begin{subfigure}[t]{0.45\textwidth}
      \includegraphics[width=7.0cm]{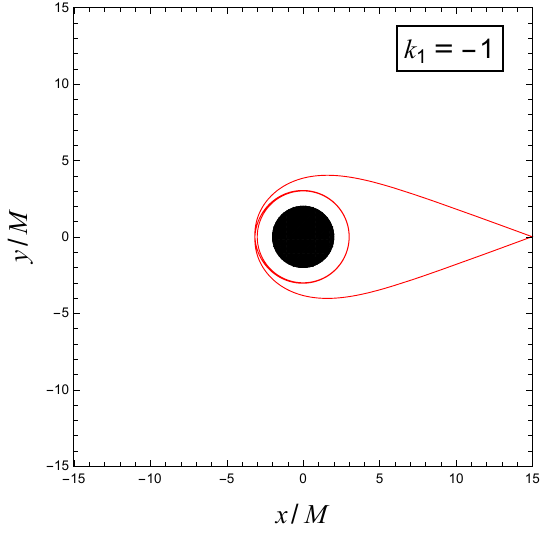}
  \end{subfigure}  
    \caption{Massive body paths starting at $x_0=15M, y_0=0$, with $ \mathscr{E}=10$ and $L=51.8747440595 M$ (we set $G=1$).  The particle either is drawn into the black hole or departs  from it,   depending on the sign of $k_1$. Setting $\LP/\RS=10^{-4}$, we have  $\R _{\rm UCO} ^{k_1=1} = 1.50168\RS$, $\R _{\rm UCO} ^{k_1=-1} = 1.50171\RS$, and $V_{\rm t} (\R _{\rm UCO} ^{k_1=1})<\mathscr{E}<V_{\rm t} (\R _{\rm UCO} ^{k_1=-1})$.}
    \label{fig:trajectorytimelike3}
\end{figure}

In Figs. \ref{fig:trajectorytimelike2} and \ref{fig:trajectorytimelike3},  we draw the trajectories resulting from Eq. \eqref{eq:massive_particle_orbit}  in the Euclidean plane, the   coordinates $x$ and $y$ being defined in the usual way
\begin{align}\label{eq:Euclidean coordinates}
    x=r \cos \phi \,,\;\; \quad  y=r \sin \phi\,.
\end{align}
In Fig.~\ref{fig:trajectorytimelike2}, although we have considered particles starting at the same location and with equal values of $\mathscr{E}$ and $L$, we find that they exhibit opposite behaviors: the body trajectory falls into the black hole if $k_1=-1$, whereas it returns almost to its initial position if $k_1=1$. This event should not be interpreted as a hint that  the black hole with $k_1=-1$ is more attractive than the one with $k_1=1$. Indeed, in  Fig.~\ref{fig:trajectorytimelike3}, the situation is completely  reversed: the particle plunges into the black hole if $k_1=1$, whereas it gets back to the initial position if $k_1=-1$. We stress that the  only difference between these two configurations is represented by the values of $L$ and $\mathscr{E}$, the latter meeting the constraint \eqref{energy-bound}.

The interesting phenomenon presented in Figs.~\ref{fig:trajectorytimelike2} and \ref{fig:trajectorytimelike3} cannot be used as mean to identify quantum signatures arising  in the Schwarzschild geometry. Indeed, to be potentially observed, we should be able to evaluate the factor $\Delta  V_{\rm t} :=  V_{\rm t} (\R _{\rm UCO} ^{k_1>0})-V_{\rm t} (\R _{\rm UCO} ^{k_1<0}) $. Since in general $\Delta  V_{\rm t}$ involves a lengthy expression, we will take  for simplicity the cases $k_1=\pm 1$ and $\mathscr{E}=1,\,2$. For $\mathscr{E}=1$, we have from Eq. \eqref{R-plus-quantum-approx} that 
\begin{align}
    \Delta  V_{\rm t} ^{\mathscr{E}=1}= \frac{L^2 \LP ^2}{64 \RS^4}+ \OO \left(\LP ^4/\RS ^4\right)\,,
\end{align}
while for $\mathscr{E}=2$, we obtain owing to Eq. \eqref{eq:sols-quartic-combine1}
\begin{align}
    \Delta  V_{\rm t} ^{\mathscr{E}=2}&= \frac{9 \left[174 \left(5 \sqrt{7}-14\right) L^2+\left(743 \sqrt{7}+1834\right) \RS^2\right] \LP ^2}{14 \left(\sqrt{7}+2\right)^4 \RS^4}+ \OO \left(\LP ^4/\RS ^4\right)\,,\notag\\
    &\approx   \frac{5.2438\LP^2}{\RS^2} - \frac{0.1852 L^2 \LP ^2}{\RS^4} + \OO \left(\LP ^4/\RS ^4\right) \,.
\end{align}
Bearing in mind that  $\R _{\rm IUCO}<\R _{\rm UCO}<\R _{\rm ISCO} $ (cf. Eq. \eqref{R-IUCO-R-3-R-ISCO}),  we  deduce from Eq.~\eqref{L-2-and-E-timelike} that $L\sim \mathscr{E} \RS$, which allows us to  conclude that 
\begin{align}\label{eq:relation-Delta V}
   \vert  \Delta  V_{\rm t} \vert \sim {\LP ^2}/{\RS ^2}\,.
\end{align}
Now, consider the following \emph{gedanken} experiment. Suppose that we prepare a light beam at some location $r_0$ outside  the black hole, with its orbit  determined by assigning specific values of $\mathscr{E}$ and $L$. Initially, we set these initial conditions to produce light rays exhibiting the same unbounded trajectory for the two distinct quantum black hole models considered here, i.e., the one with $k_1=1$ and that with $k_1=-1$. Subsequently, while keeping    $L$ fixed and increasing $\mathscr{E}$ (or vice versa, i.e.,  fixing $\mathscr{E}$ and decreasing $L$), we investigate whether the light beam maintains the same unbounded behavior for these two quantum black holes. This iterative process continues until distinct behaviors for the light beam emerge for the two different quantum black hole frameworks, as depicted in Figs. \ref{fig:trajectorytimelike2} and \ref{fig:trajectorytimelike3}. In the final stage of this procedure, the relation \eqref{eq:relation-Delta V}   implies that the energy should be fine-tuned with extreme precision  so that
\begin{align}
   \delta \mathscr{E}  \lesssim {\LP ^2}/{\RS ^2}\,,
\end{align}
or, equivalently, that  the angular momentum should be measured with  high accuracy, as 
\begin{align}
    \delta L  \lesssim {\LP ^2}/{\RS } \ll \LP\,,
\end{align}
where the $\delta$ symbol refers to the difference in either $\mathscr{E}$ or $L$ in the two scenarios  $k_1=\pm 1$ mentioned  above. Therefore,  the search for  possible  detectable properties of the quantum Schwarzschild black hole via the dynamical behaviour displayed in Figs.~\ref{fig:trajectorytimelike2} and \ref{fig:trajectorytimelike3} appears to be not feasible. Indeed, this would require a precision  in the measurement process of  $\mathscr{E}$ and $L$ which cannot be achieved, according to the  basic principles of quantum mechanics  \cite{Esposito2014}. A similar situation holds also for null geodesics, which will be studied in Sec.~\ref{Sec:null-geodesics}.

\subsubsection{Radial geodesics}\label{Sec:radial-geodesics}

The motion of massive objects in radial free fall can be described via the effective potential 
\begin{align}
V_{\rm t}^{\rm radial}(r) = \frac{1}{2} \left(1-\frac{\RS}{r}\right) + \frac{1}{2} k_1 \left(\frac{3}{2} - \mathscr{E}^2 \right) \frac{\RS \LP^2}{r^3}, 
\label{effective-potential_radial}
\end{align}
which can be promptly obtained from Eq.~\eqref{effective-potential-timelike} with $L=0$. In classical general relativity, where $k_1=0$, the effective potential always diverges towards negative infinity when $r $ approaches zero, meaning that all bodies in the vicinity of $r=0$ will be forced to plunge into the singularity.  On the other hand, in the quantum regime the  radial geodesics behavior near $r=0$ is, in general,  governed by the term proportional to $k_1$. If we suppose that this quantum correction is valid also near $r=0$, then  interesting phenomena can happen at small scales. In fact, depending on the value of the energy, there exist particles which will never fall into the singularity. In particular, for $k_1>0$ this circumstance concerns bodies with $\mathscr{E}^2<3/2$, while when $k_1<0$ it occurs when  $\mathscr{E}^2>3/2$ (particles with $\mathscr{E}^2=3/2$ will always be swallowed by the singularity if they happen to be close to $r=0$, like in the classical scenario).

This event can be readily explained in the model with $k_1>0$, where (similarly to the Reissner-Nordstr\"{o}m spacetime \cite{Poisson2009})  $r=0$ is a timelike hypersurface (see Eq. \eqref{sign-grr-1}), and hence it can be avoided by observers moving within the black hole (provided that $\mathscr{E}^2<3/2$, as we have just seen). Conversely, no classical arguments can account for this situation when $k_1<0$, since in this case $r=0$ is a spacelike hypersurface (see Eq. \eqref{sign-grr-4}). Therefore, we can suppose that a   particle is allowed to escape  the singularity because of some antigravity effects taking place in the neighbourhood of $r=0$ . These would entail the presence of some quantum matter fields which lead to the breach of NEC in the scenario with  $k_1<0$. Something similar  occurs also when $k_1>0$. Indeed, as set out in Ref. \cite{Battista2023}, this geometrical setup  permits the occurrence of a Penrose-like energy-extraction process which violates the Hawking area theorem and, as a consequence,  the NEC.  Therefore, as we discussed  also in Sec. \ref{Sec:Effective-EMT},   we can conclude that in our model the NEC is not respected independently of the sign assumed by  $k_1$, a result that ties in with the quantum nature of the Schwarzschild geometry \eqref{Schwarzschild_metric_quantum-standard}.

The validity of our considerations relies on the use of the EFT paradigm for scales where $r \approx 0$.  Although, as pointed out before, strictly speaking the EFT scheme should not be invoked in this regime, it nevertheless can provide precious hints and a first rough estimation of physical phenomena involving quantum gravity effects at small scales. As evidence of this argument, our analysis seems to point toward the  general expectation  that the theory of quantum gravity should be able to resolve the singularity issues plaguing classical general relativity \cite{Kuntz2019,Carballo-Rubio2019,Carballo-Rubio2024}. However, no possibility to assess the situations described in this section seems to be available, since they cannot be seen by any observer stationed outside  the black hole.

\subsection{Null geodesics}\label{Sec:null-geodesics}

It is well-known  that  massless particles follow a path coinciding with a null geodesic \cite{Wald}. Therefore, the dynamics of a photon can be described by setting $\alpha=0$ in the formula \eqref{effective-potential-general} of the quantum corrected effective potential, thus obtaining (see Fig. \ref{fig:Vn_eff})
\begin{align}
    V_{\rm n}(r) = \frac{L^2}{2r^2} \left(1-\frac{\RS}{r}\right)  -k_1 \frac{\mathscr{E}^2 \RS \LP^2}{2 r^3}\,.
\label{effective-potential-null}
\end{align}
In classical general relativity, the shape of the \qm{null effective potential} is independent of $L$. On the other hand, in the quantum setting $ V_{\rm n}(r)$ receives a quantum contribution  involving  the energy $\mathscr{E}$. Despite that, we will see that the formulas of the quantum  photon sphere radii  involve  neither $\mathscr{E}$ nor $L$. This result, which has a  fundamental physical significance,  will be  proved in Sec. \ref{Sec:photon-sphere}. Then, we conclude the section with the investigation of photon trajectories, see Sec. \ref{sec:Photon trajectory}. 

Similarly to the timelike geodesics case (see Eqs. \eqref{energy-bound} and \eqref{energy-bound-3}), our forthcoming analysis will performed by taking into account the upper bound $\mathscr{E} \ll \RS/\LP$, which  guarantees that quantum corrections to physical quantities never become  as important as the classical ones, in agreement with the spirit of the EFT formalism.

\subsubsection{The photon sphere}\label{Sec:photon-sphere}

The photon sphere identifies the spacetime region  where (unstable) circular light ray orbits are allowed. Its radius $ \RPS$ can be thus identified by the maximum of the effective potential \eqref{effective-potential-null},  i.e., 
\begin{align}\label{eq:V_max}
  V^\prime_{\rm n}(\RPS)=0\,,
  \end{align}
which gives the expression
\begin{align}\label{eq:r-sphere-0}
\R_{\rm ps}=\frac{3\RS}{2}+\frac{3}{2}k_1  \frac{\mathscr{E}^2 \RS \LP^2}{L^2 }\,.
\end{align}
The dependence on the ratio $\mathscr{E}/L$ can be easily removed by resorting to  the well-known identity (cf. Eq.  \eqref{geod-timelike-one-loop})
 \begin{align}
V_{\rm n}(\R_{\rm ps}) =\frac{\mathscr{E}^2}{2}\,,
\end{align} 
which yields the quantum corrected formula
\begin{align}\label{eq:r-sphere-bcr0}
\frac{\mathscr{E} ^2}{L^2}=\frac{\R_{\rm ps}-\RS}{\R_{\rm ps}^3+k_1 \RS \LP^2} \ .
\end{align}
Plugging the above relation  into Eq.~\eqref{eq:r-sphere-0}, we find the fourth-order algebraic equation
\begin{align}\label{eq:r-sphere-1}
    \R_{\rm ps}^4-\frac{3\RS}{2}\R_{\rm ps}^3 -\frac{3}{2}k_1 \RS \LP^2 \R_{\rm ps}+\frac{3}{2} k_1 \RS ^2 \LP^2=0\,,
\end{align}
whose coefficients are independent of both the energy and  angular momentum. For $k_1=0$, Eq.  \eqref{eq:r-sphere-1} reduces to a linear equation providing the single classical root
\begin{align}
\bar{\mathcal{R}}_{\rm ps 4}=\frac{3\RS}{2} \,.
\label{classical-photon-sphere}
\end{align}

In the quantum regime having $k_1\neq0$, we can find the expanded version of the roots of the quartic  \eqref{eq:r-sphere-1} by applying the analytic method devised in Sec. \ref{Sec:Timelike-geod} for the examination of  timelike geodesics. In this way, we get the following solutions: 
\begin{subequations}\label{eq:sols-quartic-null}
\begin{align}
\mathcal{R}_{\rm ps 1}&=\sqrt[3]{k_1} \RS \left(\LP /\RS \right)^{2/3}-\frac{{k_1}^{2/3}\RS}{9} \left(\LP /\RS \right)^{4/3}-\frac{2k_1 \LP^2}{27\RS}+\OO \left(\LP ^{8/3}/\RS ^{5/3}\right)\,,
\label{R-ps-quantum-1}\\
\mathcal{R}_{\rm ps 2}&=-\sqrt[3]{-k_1 }\RS\left(\LP /\RS \right)^{2/3}+\frac{{(k_1)}^{2/3}\RS}{9 (-1)^{1/3}} \left(\LP /\RS \right)^{4/3} -\frac{2k_1 \LP^2}{27\RS} +\OO \left(\LP ^{8/3}/\RS ^{5/3}\right)\,,
\\
\mathcal{R}_{\rm ps 3}&=(-1)^{2/3}\sqrt[3]{k_1 } \RS \left(\LP /\RS \right)^{2/3}-\frac{{k_1}^{2/3}\RS}{9 (-1)^{2/3}} \left(\LP /\RS \right)^{4/3}-\frac{2k_1 \LP^2}{27\RS} +\OO \left(\LP ^{8/3}/\RS ^{5/3}\right)\,,
\label{R-ps-quantum-3}\\
\mathcal{R}_{\rm ps 4}&= \bar{\mathcal{R}}_{\rm ps 4}+\frac{2k_1 \LP^2}{9\RS}+\OO \left(\LP ^{8/3}/\RS ^{5/3}\right),
\label{R-ps-quantum-4}
\end{align}
\end{subequations}
where we note that $\R_{\rm ps 1}$, $\R_{\rm ps 2}$, and $\R_{\rm ps 3}$ are fully quantum radii, consistently with the outcome of the classical analysis.  It is easy to  show that, up to  higher-order terms  in $\LP$, the Vi\`ete formulas for the quartic equations are satisfied. Thus, not surprisingly, our analytic formulas \eqref{eq:sols-quartic-null} match closely with the numerical solutions, as witnessed by  Tables \ref{table3} and \ref{table4}. 

Bearing in mind Eqs. \eqref{effective-potential-null} and \eqref{R-ps-quantum-4}, we find 
\begin{align}
V^{\prime\prime}_{\rm n}(\R_{\rm ps4})/L^2=-\left[9 \RS^2 \left(1+4 k_1 \LP^2/b^2\right)-4 k_1 \LP^2 \right]\frac{314928 \RS^4 }{\left(4 k_1 \LP^2+27 \RS^2\right)^5}+ \OO \left(\LP^{8/3}/\RS^{20/3}\right) \,,
\end{align}
which is obviously negative in light of Eqs.  \eqref{k1-O1} and \eqref{Rs-bigger-than-lp}. This means that $\R_{\rm ps4}$ is a maximum of the effective potential, and hence represents the radius of an UCO.

In principle, the roots of Eq.~\eqref{eq:r-sphere-1} can be: (i) all real, (ii) two real and two complex, (iii) all complex.   The latter possibility is naturally ruled out, since  the quartic  \eqref{eq:r-sphere-1} always admits at least one real-valued solution, i.e., the radius $\mathcal{R}_{\rm ps 4}$, which is the only one surviving in the classical limit. In addition, also the situation (i) is excluded. This is due to the fact that in Eqs. \eqref{R-ps-quantum-1}-\eqref{R-ps-quantum-3}  the coefficient $c_1$ of the leading terms (i.e., those proportional to $\LP ^{2/3}/\RS ^{2/3}$) comes from the solutions of the equation $(c_1)^3=k_1$, while the coefficient $c_2$ of the sub-leading corrections (which go like $\LP ^{4/3}/\RS ^{4/3}$) is given by the solutions of $(c_2)^3=-k_1/(9 c_1)$. If we recall that all nonzero real numbers have exactly one real cube root and a pair of complex conjugate cube roots, we soon realize that the case (i) is not present. Therefore,  only the option (ii) is possible, as is clear also from Tables \ref{table3} and \ref{table4}.

\begin{table}
    \begin{tabular}{|c|c|c|c|c|}
     \multicolumn{5}{c}{$k_1=1\,, \LP/\RS=10^{-1}$} \\
     \hhline{|=====|}
    Solution  &$\R_{\rm ps 1} /\RS$ &$\R_{\rm ps 2} /\RS $& $\R_{\rm ps 3} /\RS$  &$\R_{\rm ps 4} /\RS$  \\
    \hline
    Analytic solution&0.20955 &-0.10588-0.19105i&-0.10588+0.19105i & 1.50222 \\
    \hline
    Numerical solution&0.20945 &-0.10584-0.19098i&-0.10584+0.19098i&1.50222  \\
    \hline
    \end{tabular}
    \\ \vspace{2mm}
    \begin{tabular}{|c|c|c|c|c|}
     \multicolumn{5}{c}{$k_1=1\,, \LP/\RS=10^{-2}$} \\
     \hhline{|=====|}
Solution  &$\R_{\rm ps 1} /\RS$ &$\R_{\rm ps 2} /\RS $& $\R_{\rm ps 3} /\RS$  &$\R_{\rm ps 4} /\RS$  \\
    \hline
    Analytic solution&0.04617 &-0.02300-0.04041 i&-0.02300+0.04041i & 1.50002 \\
    \hline
    Numerical solution&0.04617 &-0.02300-0.04041i&-0.02300+0.04041i & 1.50002   \\
    \hline
    \end{tabular}
     \\ \vspace{2mm}
    \begin{tabular}{|c|c|c|c|c|}
     \multicolumn{5}{c}{$k_1=1\,, \LP/\RS=10^{-3}$} \\
     \hhline{|=====|}
Solution  &$\R_{\rm ps 1} /\RS$ &$\R_{\rm ps 2} /\RS $& $\R_{\rm ps 3} /\RS$  &$\R_{\rm ps 4} /\RS$  \\
    \hline
    Analytic solution&0.00999 &-0.00400-0.00867i&-0.00400+0.00867i & 1.50000 \\
    \hline
    Numerical solution&0.00999 &-0.00400-0.00867i&-0.00400+0.00867i & 1.50000   \\
    \hline
    \end{tabular}
     \\ \vspace{2mm}
    \begin{tabular}{|c|c|c|c|c|}
     \multicolumn{5}{c}{$k_1=1\,, \LP/\RS=10^{-4}$} \\
     \hhline{|=====|}
Solution  &$\R_{\rm ps 1} /\RS$ &$\R_{\rm ps 2} /\RS $& $\R_{\rm ps 3} /\RS$  &$\R_{\rm ps 4} /\RS$  \\
    \hline
    Analytic solution&0.00215 &-0.00108-0.00187i&-0.00108+0.00187i & 1.50000 \\
    \hline
    Numerical solution&0.00215 &-0.00108-0.00187i&-0.00108+0.00187i & 1.50000   \\
    \hline
    \end{tabular}
    \caption{Roots (in units of $\RS$) of  the quartic \eqref{eq:r-sphere-1} obtained analytically  and numerically for  $k_1=1$ and various values of $\LP/\RS$. The results provided by our analytical approach align closely with the numerical findings.}
    \label{table3}
\end{table}

\begin{table}
    \begin{tabular}{|c|c|c|c|c|}
     \multicolumn{5}{c}{$k_1=-1\,, \LP/\RS=10^{-1}$} \\
     \hhline{|=====|}
    Solution  &$\R_{\rm ps 1} /\RS$ &$\R_{\rm ps 2} /\RS $& $\R_{\rm ps 3} /\RS$  &$\R_{\rm ps 4} /\RS$  \\
    \hline
    Analytic solution&0.11104+0.18211i &0.11104-0.18211i&-0.21986 & 1.49778 \\
    \hline
    Numerical solution&0.11108+0.18220i &0.11108+0.18220i&-0.21994&1.49778  \\
    \hline
    \end{tabular}
    \\ \vspace{2mm}
    \begin{tabular}{|c|c|c|c|c|}
     \multicolumn{5}{c}{$k_1=-1\,, \LP/\RS=10^{-2}$} \\
     \hhline{|=====|}
Solution  &$\R_{\rm ps 1} /\RS$ &$\R_{\rm ps 2} /\RS $& $\R_{\rm ps 3} /\RS$  &$\R_{\rm ps 4} /\RS$  \\
    \hline
    Analytic solution&0.02334+0.03999i&0.02334-0.03999i&-0.04665 & 1.49998 \\
    \hline
    Numerical solution&0.02334+0.03999i&0.02334-0.03999i&-0.04665 & 1.49998    \\
    \hline
    \end{tabular}
        \\ \vspace{2mm}
    \begin{tabular}{|c|c|c|c|c|}
     \multicolumn{5}{c}{$k_1=-1\,, \LP/\RS=10^{-3}$} \\
     \hhline{|=====|}
Solution  &$\R_{\rm ps 1} /\RS$ &$\R_{\rm ps 2} /\RS $& $\R_{\rm ps 3} /\RS$  &$\R_{\rm ps 4} /\RS$  \\
    \hline
    Analytic solution&0.00501+0.00865i&0.00501-0.00865i&-0.01001 & 1.50000 \\
    \hline
    Numerical solution&0.00501+0.00865i&0.00501-0.00865i&-0.01001 & 1.50000  \\
    \hline
    \end{tabular}
            \\ \vspace{2mm}
    \begin{tabular}{|c|c|c|c|c|}
     \multicolumn{5}{c}{$k_1=-1\,, \LP/\RS=10^{-4}$} \\
     \hhline{|=====|}
Solution  &$\R_{\rm ps 1} /\RS$ &$\R_{\rm ps 2} /\RS $& $\R_{\rm ps 3} /\RS$  &$\R_{\rm ps 4} /\RS$  \\
    \hline
    Analytic solution&0.00108+0.00187i&0.00108-0.00187i&-0.00215 & 1.50000 \\
    \hline
    Numerical solution&0.00108+0.00187i&0.00108-0.00187i&-0.00215 & 1.50000 \\
    \hline
    \end{tabular}
    \caption{Roots (in units of $\RS$) of the quartic \eqref{eq:r-sphere-1}  for  $k_1=-1$ and different choices of $\LP/\RS$.  Our analytical approach and the numerics  demonstrate strong agreement.}
    \label{table4}
\end{table}

Our investigation reveals that  there is one photon sphere radius  when $k_1<0$ and two when $k_1>0$. This contrasts with  classical general relativity, which predicts the existence of only one photon sphere, given by Eq. \eqref{classical-photon-sphere}. In the quantum framework with $k_1$ positive, the radius $\R_{\rm ps 1}$ lies deep inside the black hole and adheres to the relation   $\R_{\rm ps 1}< \tilde{r}_1$ (see Eq. \eqref{solution-r-tilde-1}), which means that it is not visible to any observer situated outside the black hole. In this way, the possible scenarios where $\R_{\rm ps 1} \lesssim \LP$ will not be experimentally accessible.  On the other hand, orbits having radius  $\R_{\rm ps 4}$ are always visible, since $ \R_{\rm ps 4} > \tilde{r}_1$ (case $k_1>0$) and $ \R_{\rm ps 4} > \tilde{r}_3$  (case $k_1<0$),  cf. Eqs. \eqref{solution-r-tilde-1} and \eqref{solution-r-tilde-3}.

Contrary to the classical case, the effective potential \eqref{effective-potential-null} is nonvanishing when $L=0$. This means that for scales where $r \approx 0$ and when $k_1<0$, radial photons do not reach the singularity at $r=0$, despite its  spacelike nature. A similar situation holds also for radial timelike geodesics (although only for certain values of the energy, see Sec. \ref{Sec:radial-geodesics}). The consequences that can be drawn from this phenomenon are thus similar.

\subsubsection{Photon trajectories}\label{sec:Photon trajectory}

\begin{figure}[bht!]
    \centering
    \begin{subfigure}[t]{0.45\textwidth}
   \includegraphics[width=7.0cm]{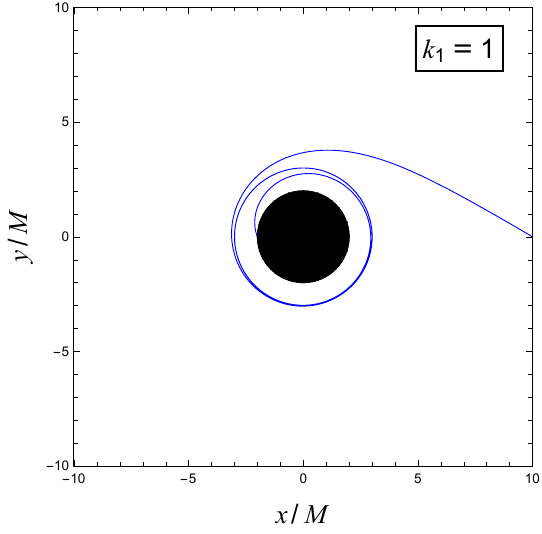}
  \end{subfigure} 
    \begin{subfigure}[t]{0.45\textwidth}
      \includegraphics[width=7.0cm]{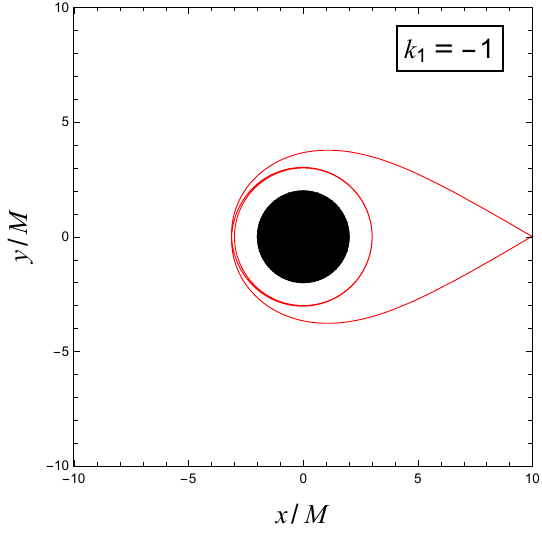}
  \end{subfigure}  
    \caption{Photon trajectories beginning at $x_0=10M, y_0=0$, with  impact parameter $b=5.19615191286M$.  Despite having identical initial conditions, two distinct orbits are obtained: the photon falls into the black hole (represented by a black disk) or returns almost to its starting point, depending on whether  $k_1=1$ or $k_1=-1$, respectively. We set $\LP/\RS=10^{-4}$ for both situations.}
    \label{fig:trajectory1}
\end{figure}

\begin{figure}[bht!]
    \centering
    \begin{subfigure}[t]{0.45\textwidth}
   \includegraphics[width=7.0cm]{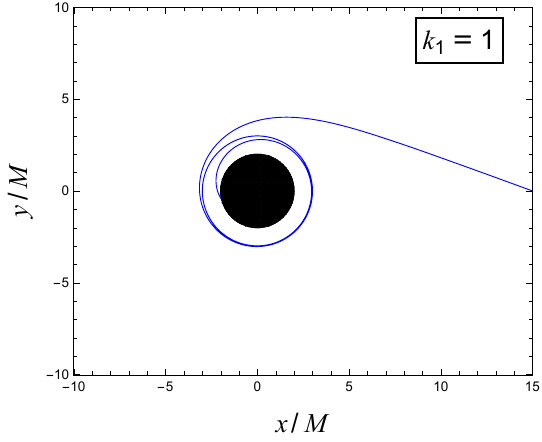}
  \end{subfigure} 
    \begin{subfigure}[t]{0.45\textwidth}
      \includegraphics[width=7.0cm]{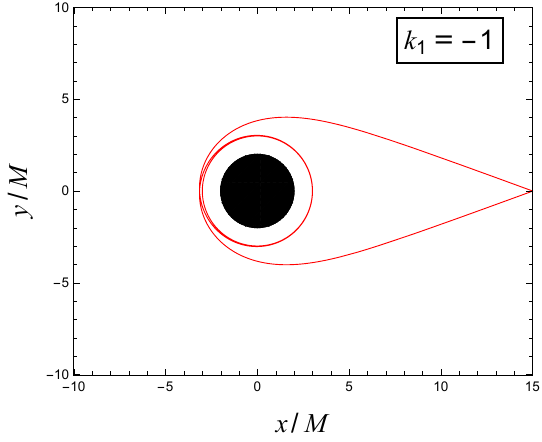}
  \end{subfigure}  
    \caption{Light ray trajectories starting at $x_0=15M, y_0=0$, with $b=5.19615156261M$. The photon either enters the black hole or moves away from it, depending on the value assumed by $k_1$. We set $\LP/\RS=10^{-4}$ for both cases. }
    \label{fig:trajectory2}
\end{figure}

\begin{figure}[bht!]
    \centering
    \begin{subfigure}[t]{1\textwidth}
   \includegraphics[width=11.0cm]{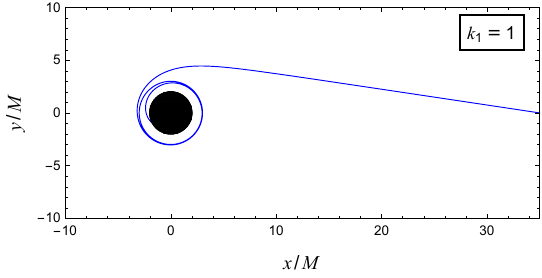}
  \end{subfigure} 
    \begin{subfigure}[t]{1\textwidth}
      \includegraphics[width=11.0cm]{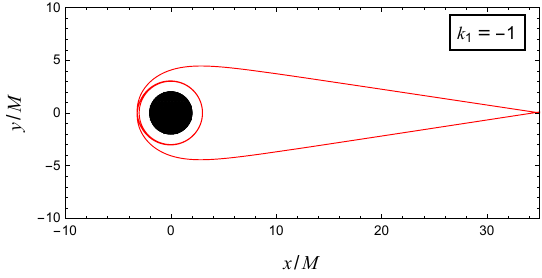}
  \end{subfigure}  
    \caption{Photon orbits with $x_0=35M$, $y_0=0$,  and $b=5.19615166247M$ showing that the motion is influenced by the sign of $k_1$. We set $\LP/\RS=10^{-4}$ for both situations.}
    \label{fig:trajectory3}
\end{figure}

It follows from Eqs. \eqref{geod-timelike-one-loop} and \eqref{effective-potential-null} that the  spatial orbits followed by light rays   on the equatorial plane $\theta =\pi/2$ of the Schwarzschild geometry \eqref{Schwarzschild_metric_quantum-standard} can be described by the equation  
\begin{align}\label{eq:photon_orbit}
    \left( \frac{\dd r }{\dd \phi }\right)^2 +\left(1-\frac{\RS}{r}\right) r^2-\frac{k_1  \RS \LP^2 \,r}{b^2}=\frac{r^4}{b^2},
\end{align}
where we have exploited the relation $\dot{\phi}= L/r^2$ and the impact parameter is still defined as $b:= L/\mathscr{E}$ (its  physical meaning will be discussed in the Sec. \ref{Sec:Shadows-rings}). The ensuing trajectories are drawn in Figs.~\ref{fig:trajectory1}-\ref{fig:trajectory3}, where we have  introduced  Euclidean coordinates according to Eq. \eqref{eq:Euclidean coordinates}. 

Quantum corrections bring forth  intriguing novel facets into photon dynamics, similar to those examined for the timelike geodesic motion, which we have  studied in Sec. \ref{Sec:trajectory-timelike}.  Let us  hereafter set $G=1$ for simplicity. In Fig.~\ref{fig:trajectory1}, we have considered a photon which starts off its orbit from the point having coordinates $x_0=10M, y_0=0$, and  with  an impact parameter  $b$  which assumes the same value for both  positive and negative  $k_1$ cases.  Since  quantum modifications to the  photon sphere radius change according to the sign assumed by $k_1$   (see Eq. \eqref{R-ps-quantum-4}), two kinds of trajectories are possible: the photon plunges into the black hole for $k_1=1$, while it approximately gets back to its initial position when $k_1=-1$. A similar phenomenon is observed also in Figs.~\ref{fig:trajectory2} and~\ref{fig:trajectory3}, albeit with different initial radii: $x_0=15 M$ and $x_0=35 M$, respectively. Therefore, our investigation shows that, although the  photon sphere radius $\R_{\rm ps 4}$ shows tiny deviations from its classical counterpart $\bar{\R}_{\rm ps 4}$, EFT theory predicts the occurrence of  completely different motions even if identical initial conditions are chosen. 
This is a remarkable result, which, nevertheless, comes with a caveat: the opposite behaviours displayed in Figs. \ref{fig:trajectory1}--\ref{fig:trajectory3} do not permit the observation of   quantum gravity signatures occurring at low-energy scales. As a consequence, the interesting features revealed in Figs. \ref{fig:trajectory1}--\ref{fig:trajectory3} cannot be employed to distinguish between frameworks with negative and positive $k_1$.  In fact,  it follows from Eqs.~\eqref{eq:r-sphere-bcr0} and \eqref{R-ps-quantum-4} that the impact parameter for the photon sphere at $r=\R_{\rm ps 4}$ is 
\begin{align}
b_{\rm ps4} =  \frac{3\sqrt{3}}{2}\RS +\frac{2k_1 \LP^2}{3\sqrt{3}\RS} + \OO \left(\LP^4/\RS^3\right) \,,
\end{align}
which means that if
\begin{align}
    b_{\rm ps 4} ^{k_1<0} \lesssim b \lesssim  b_{\rm ps 4} ^{k_1>0},
\end{align}
then it would be possible to single out distinct photon trajectories on the basis of the sign of $k_1$. However, such  accuracy cannot be achieved, because it would require that the impact parameter be adjusted  with  extreme precision, as 
\begin{align}
    b_{\rm ps 4}^{k_1>0}- b_{\rm ps 4}^{k_1<0} \lesssim \LP^2/\RS\,,
\end{align}
where we have used Eq. \eqref{k1-O1}. Physically, this is  not possible, as we have learned from quantum mechanics tenets \cite{Esposito2014}. However, while the net effect is far too small to be seen experimentally, we have found  an interesting phenomenon in its own right.

At this stage, a final consideration is in order. Our analysis  reveals an intriguing difference  with massive particles dynamics. Indeed, the photon orbits displayed in Figs. \ref{fig:trajectory1}--\ref{fig:trajectory3} exhibit a trend similar to that depicted in Fig. \ref{fig:trajectorytimelike3}, but not to the one occurring in Fig. \ref{fig:trajectorytimelike2}. In particular,  null geodesics  having $k_1=1$ (which are sketched as blue curves) always cross the (outer)  horizon, while the corresponding timelike curves  can either depart from the black hole  (Fig. \ref{fig:trajectorytimelike2}) or move into it (Fig. \ref{fig:trajectorytimelike3}).

\section{Black hole shadows and rings}\label{Sec:Shadows-rings}

The results provided in the previous section regarding the dynamics of  both massive and massless particles have a wide range of applications. In particular, they enable the  examination of the
observational  properties of the quantum black hole. 

The  landmark images  of the supermassive black holes M$87^\star$ and Sgr A$^\star$, recently released by  the Event Horizon Telescope collaboration (EHT), have marked a significant milestone in gravitational physics \cite{EventHorizonTelescope:2019dse,EventHorizonTelescope:2019ggy,EventHorizonTelescope:2022wkp}. This breakthrough achievement has put forth  the notions of  shadow
and photon ring, which, due  to their ability to provide crucial horizon-scale data, are used to  test  fundamental principles of gravity models in extreme  environments and permit inferring key black hole information such as its mass, spin, and even its structure \cite{Gralla:2019xty,Narayan:2019imo,Gralla:2020srx,Sui:2023yay,Jiang2023,Ye2023,You2024}.

Black hole shadow corresponds to the central dark area revealed by the EHT images, whose theoretical size is intricately tied to the characteristics of the emission disk. Despite that, it is commonly associated with the apparent radius of the photon sphere, which for a distant observer in standard Schwarzschild geometry is given by $b_{\rm c}=3 \sqrt{3} M$, with $b_{\rm c}$  the classical impact parameter corresponding to the classical photon sphere radius \eqref{classical-photon-sphere}. The photon ring  is a region of enhanced brightness surrounding the black hole shadow and located near  $b_{\rm c}$.

According to the analysis performed in the previous section, both the ISCO and photon sphere radii  receive quantum corrections depending on the constant $k_1$. The ensuing emission features of the black hole are unavoidably influenced by these quantum terms and are thus studied in this section. Such aspects in fact allow us  to further explore the consequences brought by the quantum modifications to the Schwarzschild geometry. After having  provided some preliminary material in Sec. \ref{Sec:Preliminaries}, the emission profiles and appearance of the black hole will be addressed in Sec. \ref{Sec:Emission-profiles}.

\subsection{Preliminaries} \label{Sec:Preliminaries}

The   emission facets  and the corresponding appearance of the black hole can be investigated by considering  the light rays trajectories  from the perspective of a distant observer, tracing them backward towards the vicinity of the black hole~\cite{Gralla:2019xty,Wang:2023rjl}. For this reason, let us  introduce the observer's local proper reference frame ($\hat{t}$, $\hat{r}$, $\hat{\theta}$, $\hat{\phi}$), which can be established by means of the non-coordinate orthonormal basis in the usual way \cite{Nakahara2003}
\begin{subequations}
\label{eq:tetrads}
\begin{align}
\hat{e}_{a}=\partial /\partial \hat{x}^{a}=e_a{} ^{\mu}\, (\partial /\partial x^{\mu}),   
\end{align}
\text{with the  dual basis satisfying}
\begin{align}
 \hat{e}^a=\dd \hat{x}^{a}=e^a{}_{\mu} \dd x^{\mu}\,,   
\end{align}
\end{subequations}
where Latin (resp. Greek) letters denote frame (resp. coordinate) indices, and $e_a{}^{\mu}$ stand for the inverse of $e^{a}{}_{\mu}$. The latter components  are commonly referred to as tetrads or vielbeins,  and form a $4\times 4$ matrix with a positive determinant  satisfying the orthonormality relation
\begin{align}
  g_{\mu \nu} =e^{a}{}_{\mu}e^{b}{}_{\nu}\,\eta_{ab}\,,
\end{align}
 $\eta_{ab}= {\rm diag}(-1,1,1,1)$ being the Lorentz metric.  When dealing with our   spherically symmetric geometry \eqref{Schwarzschild_metric_quantum-standard}, one readily obtains 
\begin{align}
    e^{a}{}_{\mu}= {\rm diag}\left( \sqrt{-g_{tt}},\sqrt{g_{rr}},\sqrt{g_{\theta \theta}},\sqrt{g_{\phi \phi}} \right) \,.
\end{align}
\begin{figure}
    \centering
    \includegraphics[scale=0.20]{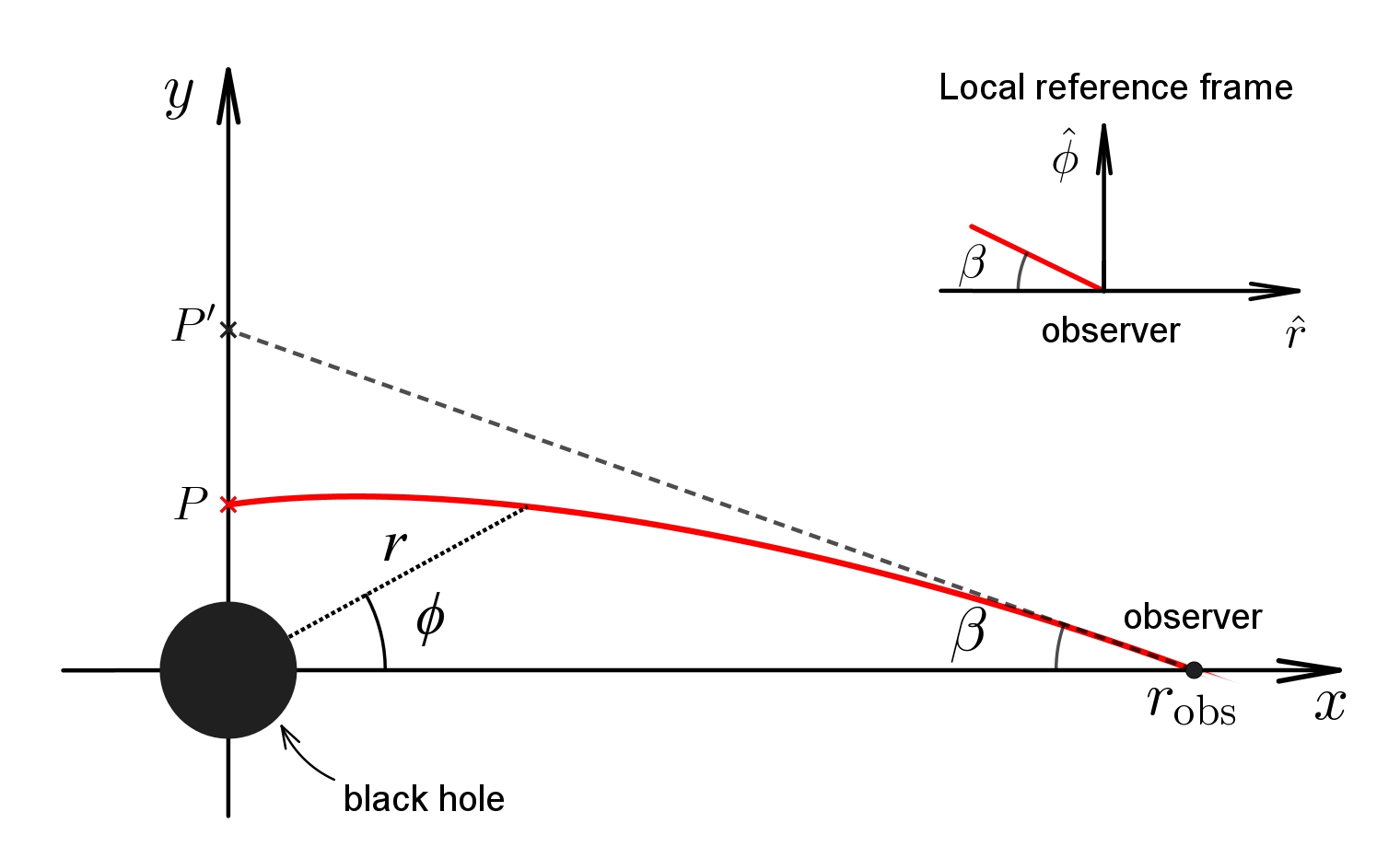}
    \caption{Photon trajectory (red curve) in the equatorial plane $\theta=\pi/2$. The main figure exhibits the Euclidean coordinates $x,y$ defined in Eq.~\eqref{eq:Euclidean coordinates}, while the inset one showcases the local reference frame of the observer positioned at radius $r_{\rm obs}$. The center of the black hole is positioned at the origin of the Euclidean plane. From the observer's perspective, light rays emitted from the point $P$ seem to originate from point $P'$.}
    \label{fig:sketchup_viewBH}
  \end{figure}
  
In Fig.~\ref{fig:sketchup_viewBH}, we depict a typical path followed by a photon as seen by an observer located at the radial distance  $r_{\rm obs}$.
When viewed from the observer's standpoint, light emitted from the point $P$ seems to originate from a different point $P'$. As a result, the perceived distance $y_{P'}$ from the center of the black hole differs from $r_{\rm obs}$ by a correcting factor, and  reads as
  \begin{align}
    y_{P'} = r_{\rm obs} \tan \beta\,,
  \end{align} 
where the angular radius $\beta$ is evaluated in the local frame as
\begin{align}\label{eq:local1}
    \tan \beta =-\dd \hat {\phi}/ \dd \hat{r}\,.
\end{align}
By means of the relation \eqref{eq:tetrads}, one can write  
\begin{align}\label{eq:local2}
    \frac{\dd \hat {\phi}}{\dd \hat{r}}=\frac{r_{\rm obs}}{\sqrt{A(r_{\rm obs})}} \frac{\dd \phi}{\dd r}\bigg| _{r_{\rm obs}}\,, 
\end{align}
and hence Eq. \eqref{eq:photon_orbit}, jointly with   formulas~\eqref{eq:local1} and ~\eqref{eq:local2},  permits to obtain  
\begin{align}
    \tan ^2 \beta =  \frac{r_{\rm obs}^2 }{{A(r_{\rm obs})}} \, \left[\frac{r_{\rm obs}^4}{b^2}-\left(1-\frac{\RS}{r_{\rm obs}}\right) r_{\rm obs}^2+\frac{k_1  \RS \LP^2 \,r_{\rm obs}}{b^2} \right]^{-1}\,,
\end{align}
 which in turn yields
\begin{align}
    \sin^2  \beta=\frac{b^2 (r_{\rm obs}-\RS)}{r_{\rm obs}^3\left[1+\OO(\LP^2/r_{\rm obs} ^2) \right]}\,,
\end{align}
where we have employed Eq.~\eqref{A-of-r-quantum}.
For a distant observer,  $r_{\rm obs} \gg \RS$ and  $\beta\ll 1$, and the above relations can be approximated as 
\begin{align}
      \tan \beta  \simeq \sin \beta \simeq b/r_{\rm obs}\,,
\end{align}
which means that 
\begin{align}
     y_{P'} \approx b := L/\mathscr{E} \,,
\end{align}
thus confirming that the factor $b$, first defined in Sec. \ref{sec:Photon trajectory}, serves as the impact parameter for null geodesics reaching the observer at infinity. 

In the subsequent analysis,  $b$ will be considered  as the apparent distance of optical sources from the center of the black hole and we will assume that the observer is situated at $r_{\rm obs}=10^{5}M$ with azimuthal angle $\phi =0$.

\subsection{Emission intensity profiles and black hole appearance}\label{Sec:Emission-profiles}

We are now ready to investigate simple scenarios where black hole emission originates from an optically and geometrically thin static disk located in its vicinity. The disk is observed face-on, and its specific intensity, denoted as $I_\nu$ (with $\nu$ representing the frequency of the emitted light in a static frame), depends solely on the radial coordinate $r$. The specific intensity emitted from the accretion disk is denoted as $I_{\rm em}(r,\nu)$. As photons are emitted from the disk, the invariant intensity, $\mathscr{I}_{\nu} := I_{\nu}/ \nu ^3$, remains constant along their trajectories in our framework, as all absorption mechanisms are neglected \cite{Gralla:2019xty,Wang:2023rjl}. Therefore, the specific intensity $I_{\rm obs} (r_{\rm obs},\nu_{\rm obs})$ received by the observer satisfies
\begin{align}\label{eq:obs_I}
   \frac{ I _{\rm em} (r,\nu)}{I _{\rm obs} (r_{\rm obs},\nu_{\rm obs})}=\left(\frac{ \nu }{\nu_{\rm obs} }\right)^3 =\left[\frac{\mathcal{G}(r_{\rm obs})}{\mathcal{G}(r)}\right]^3\,,
\end{align}  
where $\mathcal{G}(r):=\sqrt{B(r)}$ is the redshift factor (see Eq. \eqref{B-of-r-quantum}).  For a distant observer,  $r_{\rm obs} \gg \RS$, and hence we have $\mathcal{G}(r_{\rm obs}) \simeq 1$. Therefore, Eq.~\eqref{eq:obs_I} boils down to 
\begin{align}
    I_{\rm obs} (r_{\rm obs},\nu_{\rm obs})=\mathcal{G}^3 I _{\rm em} (r,\nu)\,,
\end{align}
which implies that the total observed intensity resulting from light rays emitted from a specific location $r$ can be integrated as 
\begin{align}
    I_{\rm obs} (r_{\rm obs}) = \int I_{\rm obs} (r_{\rm obs},\nu_{\rm obs}) \dd \nu_{\rm obs}=\int  \mathcal{G}^4 I _{\rm em} (r,\nu) \dd \nu= \mathcal{G}^4 I_{\rm em}(r)\,,
\end{align}
where $I_{\rm em}(r) \equiv \int   I _{\rm em} (r,\nu) \dd \nu $ is the integrated intensity. 

When tracing a light ray backward from the observer, there exists the possibility of intersecting the accretion disk, leading to an increase in brightness due to the disk emission. The frequency of such intersections determines the accumulated brightness along the ray's path. The total observed intensity is derived by summing the intensities contributed by each intersection~\cite{Gralla:2019xty} 
\begin{align}
I_{\rm obs} (b)=\sum_{m}\left[\mathcal{G}^4 I_{\rm em}\right]\Big|_{r=r_m(b)}\,,
\label{obs-intensity} 
\end{align}
where $r_m(b)$ ($m=1,2,3,\dots$) denotes the radial coordinate of the $m$-th intersection position with the disk plane outside the black hole. For subsequent numerical computations, we will consider only the first three intersections.

Our investigation of the emission intensity profiles relies on the  Gralla-Lupsasca-Marrone (GLM) model~\cite{Gralla:2020srx}, which yields predictions closely aligned with those from general relativistic magneto-hydrodynamics simulations of astrophysical accretion disks~\cite{Gralla:2020srx,Rosa:2023qcv}. The emission intensity profile of the GLM pattern is given by~\cite{Gralla:2020srx}
\begin{align}\label{eq:GLM}
I_{\rm em}(r)=\frac{\rm{e}^{-\frac{1}{2}\left[\gamma +\text{arcsinh} (\frac{r-\mu}{\sigma})\right]^2}}{\sqrt{(r-\mu)^2+\sigma ^2}}\,,
\end{align}
where $\mu$, $\gamma$, and $\sigma$ are free phenomenological parameters  governing the  emission shape. 
\begin{figure}[bht!]
\centering
\begin{subfigure}[t]{0.45\textwidth}
\includegraphics[width=7.5cm]{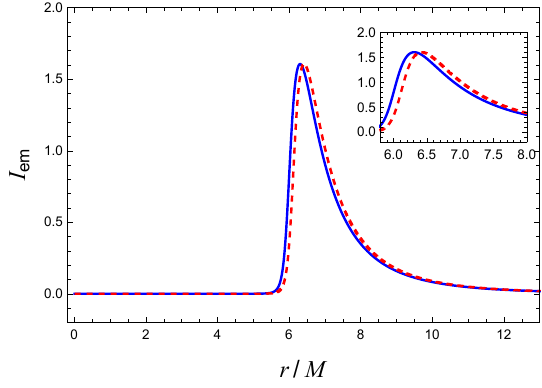}
\caption{ISCO model}
\label{fig:shadow1_emission}
\end{subfigure} 
\begin{subfigure}[t]{0.45\textwidth}
\includegraphics[width=7.5cm]{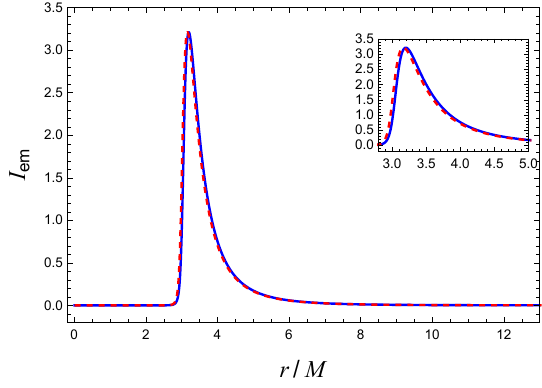}
\caption{PS model}
\label{fig:shadow1_emission2}
\end{subfigure}  
\caption{ISCO and PS models given by Eq.~\eqref{eq:GLM}.  The blue and red dashed curves represent the emission intensity profiles for the black hole with  $k_1=1$ and $k_1=-1$, respectively. $\mathcal{R}_{\rm ISCO}$ and $\mathcal{R}_{\rm ps 4}$ have been evaluated  by taking $\LP/\RS =0.2$. The small boxes show zoomed-in details of the plots.  }
\label{Fig:emission-profiles}
\end{figure}
\begin{figure}[bht!]
\centering
\begin{subfigure}[t]{0.45\textwidth}
\includegraphics[width=7.5cm]{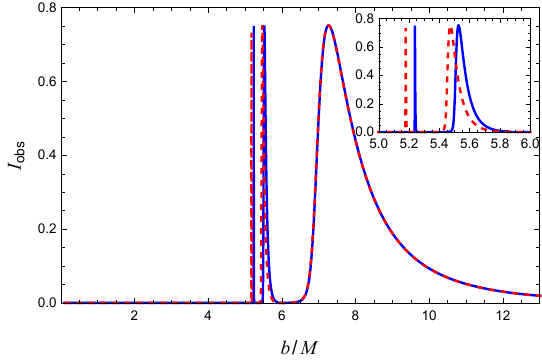}
\caption{$\LP/\RS =0.2$}      
\end{subfigure} 
 \begin{subfigure}[t]{0.45\textwidth}
\includegraphics[width=7.5cm]{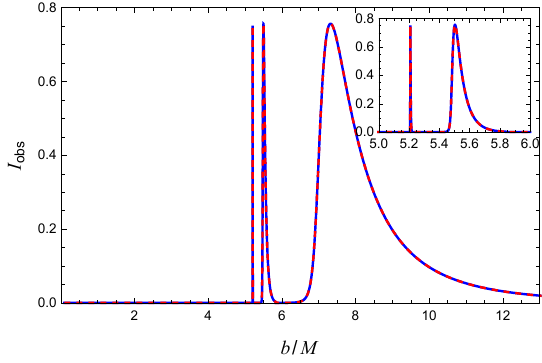}
\caption{$\LP/\RS =10^{-4}$}      
\end{subfigure} 
 \caption{Observed intensity \eqref{obs-intensity} of the ISCO model for different values of the ratio $\LP/\RS$. The blue and red dashed curves represent the  profile for the black hole with  $k_1=1$ and $k_1=-1$, respectively.  The inset figures depict zoomed-in  portions of the plots. }
    \label{fig:observerI1compare}
\end{figure}
\begin{figure}[bht!]
    \centering
\begin{subfigure}[t]{0.45\textwidth}
\includegraphics[width=7.5cm]{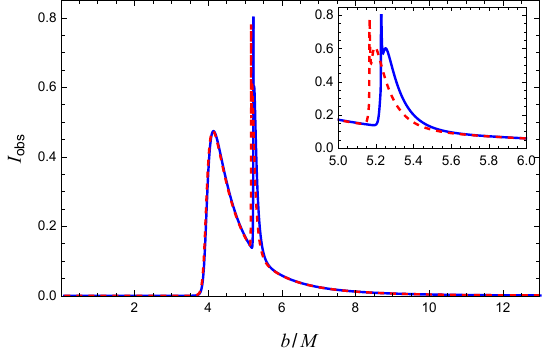}
\caption{$\LP/\RS =0.2$}      
\end{subfigure} 
\begin{subfigure}[t]{0.45\textwidth}
\includegraphics[width=7.5cm]{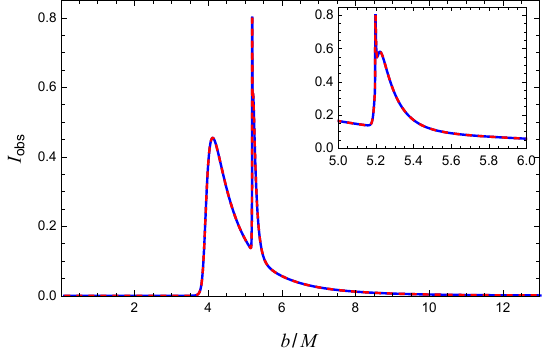}
\caption{$\LP/\RS =10^{-4}$}
\end{subfigure} 
\caption{Observed intensity \eqref{obs-intensity} of the PS model for different values of the ratio $\LP/\RS$. The blue and red dashed curves represent the  profile for the black hole with  $k_1=1$ and $k_1=-1$, respectively.  Small boxes within the figures display zoomed-in parts of the plots. }
\label{fig:observerI1compare2}
\end{figure}
\begin{figure}[bht!]
 \centering
    \begin{subfigure}[t]{0.4\textwidth}
       \includegraphics[width=6.0cm]{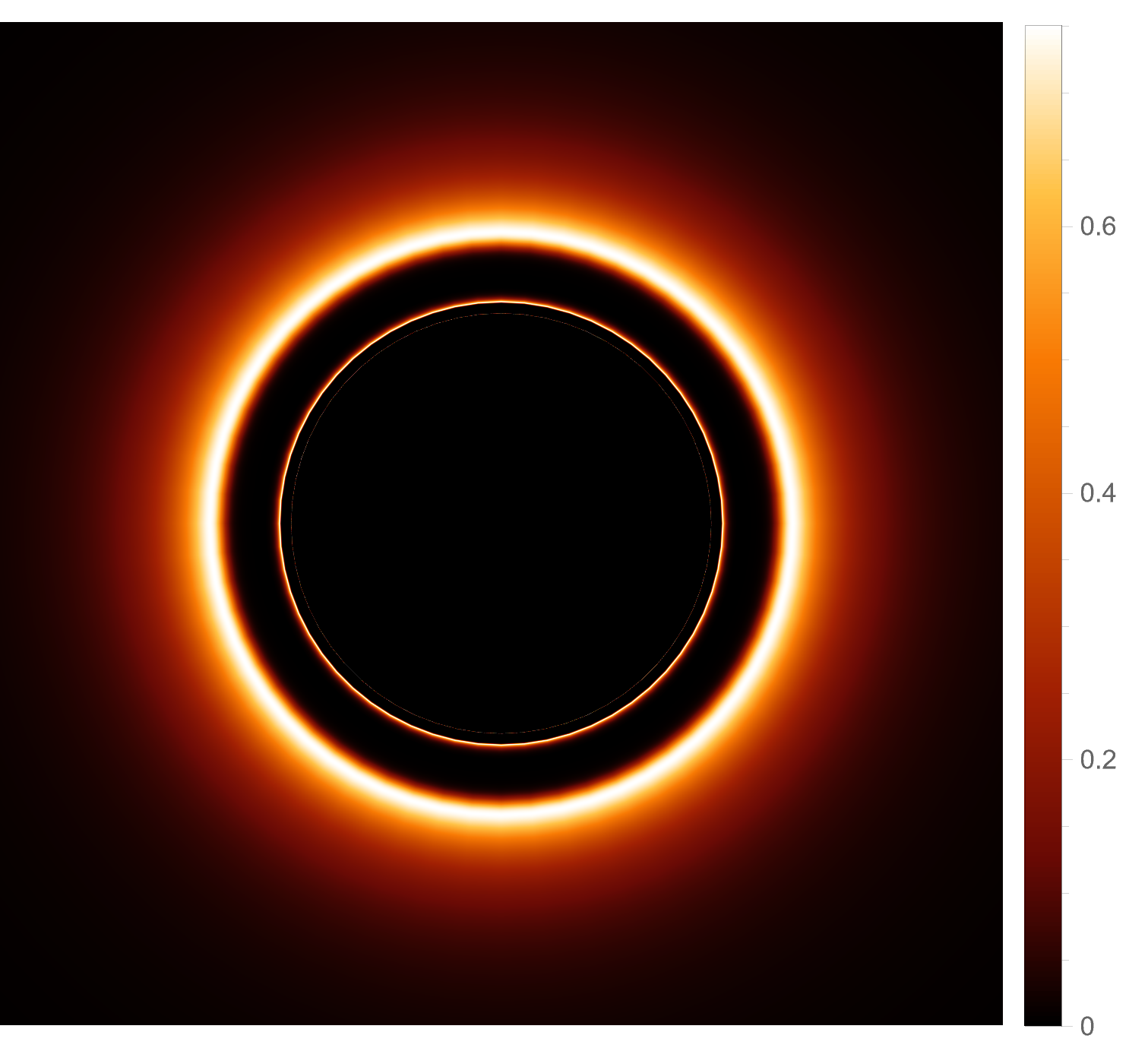}
       \caption{ISCO model}
  \end{subfigure} 
  \begin{subfigure}[t]{0.4\textwidth}
      \includegraphics[width=6.0cm]{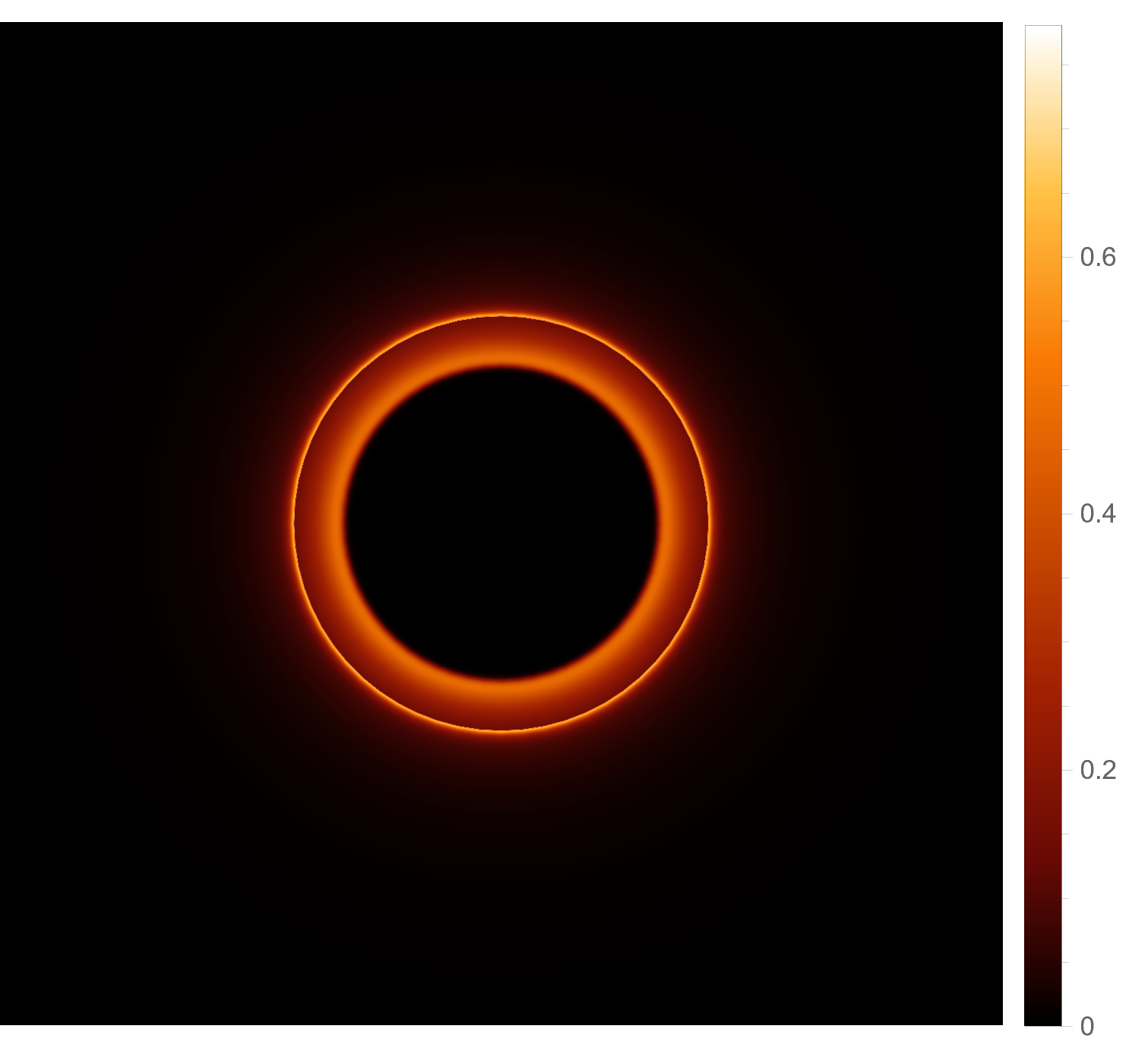}
      \caption{PS model}
       \end{subfigure} 
\caption{Observational appearance for the ISCO and PS models of an optically thin disk of emission. We choose $\LP/\RS =0.2$, and  $k_1=1$ for left panel and $k_1=-1$ for the right panel. Information regarding the observed intensity $I_{\rm obs}$ is reported on the right side of the figures. In the ISCO model, the photon ring appears as the thin circle within the dark area, while in the PS one, it appears as the outer ring surrounding the dark region. }
  \label{fig:two_figures}
\end{figure}
We will consider the following two different GLM profiles  \cite{Gralla:2020srx,Rosa:2023qcv,Rosa:2023hfm}:
\begin{itemize}
\item ISCO model: the emission graph is broadly peaked at the quantum radius $r= \mathcal{R}_{\rm ISCO}$ (see Eq. \eqref{R-ISCO-new-method});  minimal emission occurs when $r< \mathcal{R}_{\rm ISCO}$, since  in this region  only UCOs are allowed, as we have shown previously. Following Ref. \cite{Gralla:2020srx}, for this framework we  choose $\mu= \mathcal{R}_{\rm ISCO}$, $\gamma =-2$, and $\sigma =M/4$. 
\item Photon sphere (PS) model: the emission curve peaks at the quantum position $r=\mathcal{R}_{\rm ps 4}$  (see Eq. \eqref{R-ps-quantum-4}), since there exist UCOs within the region $\mathcal{R}_{\rm ps 4}<r<\mathcal{R}_{\rm ISCO}$, as we we seen before. Inspired by Ref. \cite{Gralla:2020srx}, in this pattern we take $\mu = \mathcal{R}_{\rm ps 4}$, $\gamma =-2$, and $\sigma =M/8$. 
\end{itemize}

The emission profiles and the observed appearances of the quantum Schwarzschild black hole are displayed in Figs. \ref{Fig:emission-profiles}, and \ref{fig:observerI1compare} -- \ref{fig:two_figures}, respectively. It is clear that   our analysis reveals tiny departures from the classical Schwarzschild solution, even in the limit scenario with $\LP/\RS =0.2$.  In other words, black hole shadows seem not to  provide a mean to discriminate among the possible values of the factor $k_1$ occurring in the EFT literature devoted to study of quantum black hole geometries. In that regard, we also note that, for the scopes of this section, both the choice $\LP/\RS =0.2$ and $\LP/\RS =10^{-4}$  should be regarded as very extreme situations, although they still satisfy the constraint \eqref{Rs-bigger-than-lp}. Indeed, they yield primordial black holes having a mass $M \simeq 5 \times 10^{-8} \, {\rm kg} \sim 2.5 \, \MP $  in the first case, and   $M \simeq 1 \times 10^{-4} \, {\rm kg} \sim 5 \times 10^4 \, \MP$  in the second.

\section{Concluding remarks}\label{Sec:conclusions}

The issues related to the nonrenormalizability of Einstein theory can be overcome via the EFT approach, which permits to derive the leading one-loop long-distance quantum corrections arising in gravity interactions. 

In this paper, we have analyzed the quantum Schwarzschild geometry within the EFT pattern. After having outlined its main facet along with some thermodynamic aspects in Sec. \ref{Sec:Quantum-metric}, in Sec. \ref{Sec:geodesic-motion-massive-massless} we have examined the behaviour of timelike and null geodesics.  We have developed an efficient analytic method for solving  the fourth-order algebraic equations governing the  quantum positions of SCOs and UCOs. Consistently with the EFT recipes, the obtained roots are expressed in a readable way as a power series in the ratio $\LP/\RS$, and have allowed us  to compute  the quantum version of the ISCO and PS radii (see Eqs. \eqref{eq:sols-quartic}--\eqref{eq:sols-quartic-combine2}, \eqref{R-ISCO-new-method}, and \eqref{eq:sols-quartic-null}). In this way, we have discovered that the scenario having positive $k_1$ admits two disconnected SCO regions, and, as a consequence, a second ISCO radius, as well as  two PS radii. These additional radii are located deep inside the black hole and are not visible from outside. Such characteristics  are to be added to those already known from Sec. \ref{Sec:horizon-null-hypers}, i.e., the presence of  two horizons and two null hypersurfaces where $g^{rr}=0$. These can be regarded as drawbacks of the framework with $k_1>0$ when it is compared to the classical Schwarzschild solution. 

The examination of massive and massless particles equations of motion has revealed an intriguing peculiarity of the quantum Schwarzschild black hole. Indeed, even if  identical initial conditions are chosen, the ensuing orbits exhibit different behaviours depending on the sign of $k_1$ (see Figs. \ref{fig:trajectorytimelike2}, \ref{fig:trajectorytimelike3}, and  \ref{fig:trajectory1}--\ref{fig:trajectory3}). This means that, although leading quantum contributions slightly alter the  properties of the classical Schwarzschild spacetime, they can make causal geodesics  evolve in completely different ways. Unfortunately, it appears that this phenomenon yields no direct observable signatures of the quantum corrections affecting the dynamics, and hence cannot be exploited to single out the possible values of $k_1$ occurring in the quantum Schwarzschild metric \eqref{Schwarzschild_metric_quantum-standard}.

In Sec. \ref{Sec:Shadows-rings}, we have considered a first application of the results concerning the geodesic motion by dealing with  black hole shadows and rings,  which have garnered considerable interest in the recent literature since the groundbreaking images disclosed by EHT collaboration. Similarly as before, quantum imprints in the ISCO and PS emission profiles seem to indicate no visible consequences (see Figs. \ref{Fig:emission-profiles}--\ref{fig:two_figures}).

One  interesting aspect of the quantum Schwarzschild solution is that  the   effective potential \eqref{effective-potential-general} depends explicitly  on  the energy. Such characteristic has allowed us to  work  out a cutoff energy scale for our model involving  the ratio   $\RS/\LP$, or equivalently $M/\MP$ (see Eqs. \eqref{energy-bound} and \eqref{energy-bound-3}). This ties in with the principles of EFT scheme, which entail that new particles or degrees of freedom become important at specific energy scales. Broadly speaking, energy-dependent quantum effects can have different origins and lead to significant implications. First of all,   they can hint at the need for renormalization procedures  and  indicate the running of coupling constants, which are known to be basic ingredients of a quantum field theory. Furthermore, they can also be connected with phenomena like the quantum tunneling, which  influences the behavior of particles in potential wells.  

The results obtained in this paper can trigger the study of further  relevant  topics. In particular,
a key point to be addressed concerns the search for quantum phenomena which can give rise to some measurable outcomes. This circumstance will permit to  determine the correct value of the $k_1$ factor, thereby resolving the discrepancies  in the existing literature. Our analysis seems to point toward the case with $k_1<0$, which however is plagued by the fact that it predicts no bounds on the mass $M$, while the other paradigm naturally implements the constraint \eqref{M-bigger-M-Planck} (recall in fact that the horizons \eqref{solution-r1} and \eqref{solution-r2} can be defined only if $M>M^\star \sim \MP$). These subjects, along with the investigation of other fundamental properties of the quantum Schwarzschild black hole, deserve a careful consideration in a separate paper.

\section*{Acknowledgements}

The work of Z. W. is supported by the Natural Science Foundation of Jiangsu Province (BK20220642). E. B. acknowledges the support of INFN {\it sezione di Napoli}, {\it iniziativa specifica} QGSKY and Moonlight2. E. B. thanks D-IAS for the hospitality and support. E. B. thanks Dr. R. Marotta for useful discussions and clarifications on some of the topics investigated in the paper.

\bibliographystyle{utphys}
\bibliography{references}{}

\end{document}